\documentstyle[amssymb,thmsa,leqno,sw20mitp]{article}

\makeatletter
\renewcommand\theequation{\thesection.\arabic{equation}}
\@addtoreset{equation}{section}
\makeatother
\input{tcilatex}
\begin{document}

\title{Random Partitioning Problems Involving Poisson Point Processes On The
Interval}
\author{Thierry HUILLET \\
Laboratoire de Physique Th\'{e}orique et Mod\'{e}lisation, \\
CNRS-UMR 8089 et Universit\'{e} de Cergy-Pontoise, \\
2 Avenue Adolphe Chauvin, 95032, Cergy-Pontoise, FRANCE.\\
E-mail: Thierry.Huillet@ptm.u-cergy.fr}
\maketitle

\begin{abstract}
Suppose some random resource (energy, mass or space) $\chi \geq 0$ is to be
shared at random between (possibly infinitely many) species (atoms or
fragments). Assume ${\Bbb E}\chi =\theta <\infty $ and suppose the amount of
the individual share is necessarily bounded from above by $1$. This random
partitioning model can naturally be identified with the study of infinitely
divisible random variables with L\'{e}vy measure concentrated on the interval%
$.$ Special emphasis is put on these special partitioning models in the
Poisson-Kingman class. The masses attached to the atoms of such partitions
are sorted in decreasing order. Considering nearest- neighbors spacings
yields a partition of unity which also deserves special interest. For such
partition models, various statistical questions are addressed among which:
correlation structure, cumulative energy of the first $K$ largest items,
partition function, threshold and covering statistics, weighted partition,
R\'{e}nyi's, typical and size-biased fragments size. Several physical images
are supplied.

When the unbounded L\'{e}vy measure of $\chi $ is $\theta x^{-1}\cdot {\bf I}%
\left( x\in \left( 0,1\right) \right) dx$, the spacings partition has
Griffiths-Engen-McCloskey or GEM$\left( \theta \right) $ distribution and $%
\chi $ follows Dickman distribution. The induced partition models have many
remarkable peculiarities which are outlined.

The case with finitely many (Poisson) fragments in the partition law is also
briefly addressed. Here, the L\'{e}vy measure is bounded.\newline

KEYWORDS:{\em \ }Random Partitions, Divisibility, Poisson Point Process on
the Interval.\newline

AMS 1991: Primary 60G57, 62E17, Secondary: 60K99, 62E15, 62E20 \newpage
\end{abstract}

\section{Introduction}

\indent

Random division models of a population into a (possibly large) number $n$ of
species, fragments or valleys with random weights or sizes have received
considerable attention in various domains of applications.

In disordered systems Physics, it was first recognized as an important issue
in \cite{Der}, as a result of phase space (in iterated maps or spin glasses
models at thermal equilibrium) being typically broken into many valleys, or
attraction basins, each with random weight. Problems involving the breakdown
or splitting of some item into random component parts or fragments, also
appear in many other fields of interest: for example the composition of
rocks into component compounds in Geology (splitting of mineral grains or
pebbles), the composition of biological populations into species or the
random allocation of memory in Computer Sciences, but also models for gene
frequencies in population genetics and biological diversity.

All these applications deal with randomly broken objects and random
splitting (see also \cite{Feller} pages 25 and 30 for further motivations).
Considering the random weights of the various species must sum to one, by
normalization, the typical phase-space of these models is the interval of
unit length, randomly split in such a way that the fragments' masses, sizes
or energies must sum to one. The random structure of the population is then
characterized by the ranked sequence of fragments' weights or sizes. This
was observed in \cite{Derrida} (in the large $n$ thermodynamic limit i.e.
with a denumerable number of fragments).\newline

There are of course many ways to break the interval at random into $n$
(possibly infinitely many) pieces and so one needs to be more specific. This
manuscript is precisely devoted to the study of some remarkable partition
laws of the interval which arise from the partitioning problem of some
random variable.\newline

More precisely, in Section 2, we shall first focus on the simplest ``fair''
statistical model for splitting the interval into a finite number $n$ of
fragments. It essentially relies on normalization of a sequence of random
variables by its sum. In more details, let $n>1$ be a given integer. With $%
\left( S_{k};k\geq 1\right) $ independent and identically distributed
positive random variables, consider the partial random walk sum $\chi
_{n}:=S_{1}+..+S_{n}$. Then $S_{1},...,S_{n}$ constitute a simple random
partition of $\chi _{n}$. Normalizing, define $\varsigma _{k}=S_{k}/\chi
_{n} $, $k=1,..,n.$ Then $\left( \varsigma _{1},...,\varsigma _{n}\right) $
constitutes a random partition of unity. In this model, there are $%
1<n<\infty $ fragments with exchangeable random sizes $\left( \varsigma
_{1},...,\varsigma _{n}\right) $ summing up to $1$. We shall focus on the
special case where $S_{1}$ has gamma$\left( \alpha \right) $ distribution,
with $\alpha >0$. In this case, $\left( \varsigma _{1},...,\varsigma
_{n}\right) $ has Dirichlet D$_{n}\left( \alpha \right) $ distribution. Let $%
\left( \varsigma _{\left( 1\right) },...,\varsigma _{\left( n\right)
}\right) $ be the ranked version of $\left( \varsigma _{1},...,\varsigma
_{n}\right) $, with $\varsigma _{\left( 1\right) }>...>\varsigma _{\left(
n\right) }$. Passing to the weak limit $n\uparrow \infty $, $\alpha
\downarrow 0$ with $n\alpha =\theta $, one gets the ranked Poisson-Dirichlet
partition model PD$\left( \theta \right) $. It may also be obtained from the
normalization process of the jumps of a Moran subordinator, resulting in a
random discrete distribution on the infinite simplex; see \cite{Kingman}. We
shall recall some of its remarkable properties. The PD model exhibits many
fundamental invariance properties. For a review of these results and
applications to Computer Science, Combinatorial Structures, Physics,
Biology.., see \cite{Tavare} and the references therein for example; this
model and related ones are also fundamental in Probability Theory; see \cite
{Pit1}, \cite{Pit2}, \cite{Pit3} and \cite{PitYor}.\newline

Several (not exclusively) interesting partitioning models of the interval
are based on such normalizing process in the literature. In the sequel, we
shall study different types of partitioning models rather based on
nearest-neighbor spacings.\newline

More specifically, in Section 3, we shall indeed discuss the following
closely related partitioning model. Let $\chi >0$ be an infinitely divisible
random variable with L\'{e}vy measure concentrated on $\left( 0,1\right) $
with total mass $\infty $ (the so-called unbounded case). Assume ${\Bbb E}%
\chi =\theta <\infty $. Then the partition $\chi \stackrel{d}{=}\sum_{k\geq
1}\xi _{\left( k\right) }$ is obtained from the constitutive ordered jumps $%
\xi _{\left( 1\right) }>...>\xi _{\left( k\right) }>...$ of $\chi $. The
system $\left( \xi _{\left( k\right) };k\geq 1\right) $ constitutes a
Poisson point process on $\left( 0,1\right) $. Special emphasis is put on
these partitioning models of the Poisson-Kingman type in Section 3.1. Their
specificity is that each fragment in the decomposition of $\chi $ has size
physically bounded from above by $1$. Several statistical questions arising
in this partitioning context are then discussed among which: fragments
correlation structure, cumulative sum of the $K$ largest items, partition
function, threshold statistics, filtered partition, typical and size-biased
picked fragments size. All these statistical questions are of concrete
interest.

In Section 3.2, the following related partition is also considered: let $%
\widetilde{\xi }_{k}=\xi _{\left( k-1\right) }-\xi _{\left( k\right) }$, $%
k\geq 1$, (with $\xi _{\left( 0\right) }:=1$) stand for spacings between
consecutively ordered $\xi _{\left( k\right) }$s. Then, $\sum_{k\geq 1}%
\widetilde{\xi }_{k}=1$ and $\left( \widetilde{\xi }_{k},k\geq 1\right) $
constitutes an alternative random partition of unity. In sharp contrast with
limiting partitioning models of Section 2, its construction does not involve
any normalization procedure. Its specificity rather is a consequence of the
L\'{e}vy measure for jumps of $\chi $ to be concentrated on $\left(
0,1\right) $ leading to $\left( 0,1\right) -$valued $\xi _{\left( k\right) }$%
s. Similar statistical questions arising in this partition context are also
addressed.\newline

In Section 4, a remarkable special case of the partitioning models developed
in Section 3 is studied in some detail. It corresponds to the following
particular model: assume the unbounded L\'{e}vy measure of $\chi $ takes the
particular form: $\theta /x\cdot {\bf I}_{x\in \left( 0,1\right) }dx$. Then,
the random variable $\chi $ has Dickman distribution. The induced
partitioning models have many remarkable peculiarities which are outlined
throughout. In particular, the spacings partition has
Griffiths-Engen-McCloskey or GEM$\left( \theta \right) $ distribution whose
ordered version is Poisson-Dirichlet partition. The PD$\left( \theta \right) 
$ was obtained in Section 2 from a very different construction based on
normalization.\newline

Finally, in Section 5, we assume that the L\'{e}vy measure of $\chi $ is now
with finite total mass (the bounded case). In this case, we are led to
random partitions of $\chi $ or of unity into a finite Poissonian number of
fragments. Some of their properties are briefly outlined.

\section{Exchangeable Dirichlet Partition with Finitely Many Fragments and
its Poisson-Dirichlet Limit}

\indent

We start with recalling a standard construction of the Poisson-Dirichlet
partition as a limiting partition from the exchangeable Dirichlet partition
of unity.\newline

\begin{center}
{\bf Dirichlet partition}
\end{center}

Suppose there are $\infty >n>1$ fragments with random sizes, say $\left(
\varsigma _{1},...,\varsigma _{n}\right) ,$ where $\left( \varsigma
_{1},...,\varsigma _{n}\right) $ has exchangeable distribution, implying in
particular that each $\varsigma _{k}$, $k=1,..,n$ all share the same
distribution, say the one of $\varsigma \stackrel{d}{=}\varsigma _{1}$ (each
item has statistically the same mass). We also assume that $\varsigma $ has
a density $f_{\varsigma }\left( x\right) >0$ on $\left( 0,1\right) $ with
total mass $1$ and that $\sum_{k=1}^{n}\varsigma _{k}=1$ (almost surely)
which is a strict conservativeness property of the partition.

With $\alpha >0$, we assume more specifically that $\left( \varsigma
_{1},..,\varsigma _{n}\right) $ is distributed according to the
(exchangeable) Dirichlet-D$_{n}\left( \alpha \right) $ density function on
the simplex meaning 
\begin{equation}
f\left( x_{1},..,x_{n}\right) =\frac{\Gamma \left( n\alpha \right) }{\Gamma
\left( \alpha \right) ^{n}}\prod_{k=1}^{n}x_{k}^{\alpha -1}\cdot \delta
_{\left( \sum_{k=1}^{n}x_{k}-1\right) }.  \label{eq1}
\end{equation}
In this case, $\varsigma _{k}\stackrel{d}{=}\varsigma $, $k=1,..,n$ and the
individual fractions are all identically distributed. Their common density
on the interval $\left( 0,1\right) $ is given by 
\[
f_{\varsigma }\left( x\right) =\frac{\Gamma \left( n\alpha \right) }{\Gamma
\left( \alpha \right) \Gamma \left( \left( n-1\right) \alpha \right) }%
x^{\alpha -1}\left( 1-x\right) ^{\left( n-1\right) \alpha -1}. 
\]
This is the one of a beta$\left( \alpha ,\left( n-1\right) \alpha \right) $
random variable, with moment function 
\begin{equation}
\varphi _{\varsigma }\left( q\right) :={\Bbb E}\left( \varsigma ^{q}\right) =%
\frac{\Gamma \left( n\alpha \right) }{\Gamma \left( n\alpha +q\right) }\frac{%
\Gamma \left( \alpha +q\right) }{\Gamma \left( \alpha \right) }\text{, }%
q>-\alpha .  \label{eq2}
\end{equation}

The case $\alpha =1$ corresponds to the uniform partition into $n$ fragments
for which 
\[
\varphi _{\varsigma }\left( q\right) =\frac{\left( n-1\right) !}{\left(
q+n-1\right) (q+n-2)...(q+1)}\text{, }q>-1. 
\]
This remarkable partition model is in the larger class of those for which $%
\varsigma _{k}=S_{k}/\left( S_{1}+..+S_{n}\right) $ where the $S_{k}$, $%
k=1,..,n$ are independent and identically distributed (iid) positive random
variables. Indeed, assuming $S_{1}\stackrel{d}{\sim }$ gamma$\left( \alpha
\right) $, the joint distribution of $\left( \varsigma _{1},..,\varsigma
_{n}\right) $ is well-known to be Dirichlet D$_{n}\left( \alpha \right) .$%
\newline

\begin{center}
{\bf Poisson-Dirichlet partition and the Kingman limit}
\end{center}

In such ``equitable'' Dirichlet model, consider the situation where $%
n\uparrow \infty $, $\alpha \downarrow 0$ while $n\alpha =\theta >0$. Such
an asymptotic was first considered by \cite{Kingm}. As noted by Kingman, $%
\left( \varsigma _{1},..,\varsigma _{n}\right) $ $\stackrel{d}{\sim }$ D$%
_{n}\left( \alpha \right) $ itself has no non-degenerate limit. However,
considering the ranked version $\left( \varsigma _{\left( 1\right)
},..,\varsigma _{\left( n\right) }\right) $ with $\varsigma _{\left(
1\right) }>..>\varsigma _{\left( n\right) }$, one may check that in the
Kingman limit, $\left( \varsigma _{\left( k\right) },k=1,..,n\right) $
converges in law to a Poisson-Dirichlet distribution, say $\left( \varsigma
_{\left( k\right) }\text{, }k\geq 1\right) \stackrel{d}{\sim }$ PD$\left(
\theta \right) $ with $\varsigma _{\left( 1\right) }>..>\varsigma _{\left(
k\right) }>...$ The size-biased permutation of $\left( \varsigma _{\left(
k\right) }\text{, }k\geq 1\right) $ is, say $\left( \varsigma _{k}\text{, }%
k\geq 1\right) \stackrel{d}{\sim }$ GEM$\left( \theta \right) $, the
so-called Griffiths-Engen-McCloskey law (see \cite{Kingman}, Chapter $9$).
For this partition of unity, the following Residual Allocation Model (or
RAM) decomposition holds 
\begin{equation}
\varsigma _{k}=\prod_{l=1}^{k-1}\overline{v}_{l}v_{k}\text{, }k\geq 1.
\label{eq5}
\end{equation}
Here $\left( v_{k},k\geq 1\right) $ are iid with common law $v_{1}\stackrel{d%
}{\sim }$ beta$\left( 1,\theta \right) $ and $\overline{v}_{1}:=1-v_{1}%
\stackrel{d}{\sim }$ beta$\left( \theta ,1\right) $. Note that $\varsigma
_{1}\succeq _{st}..\succeq _{st}\varsigma _{k}\succeq _{st}..$, and that $%
\left( \varsigma _{k}\text{, }k\geq 1\right) $ is invariant under
size-biased permutation.\newline

As is well-known, the Poisson-Dirichlet partition can be understood as
follows (see \cite{Kingman}, Chapter $9$ and \cite{Holstb}). Let $\chi >0$
be some infinitely divisible random variable whose L\'{e}vy measure $\Pi
\left( dx\right) $ is concentrated on $\left( 0,\infty \right) ,$ with
infinite total mass. Assume more specifically that $\Pi \left( dx\right)
=\theta e^{-x}/x,$ $x>0.$ This is the L\'{e}vy measure for jumps of a Moran
(gamma) subordinator $\left( \chi _{t};t\geq 0\right) $, with $\chi :=\chi
_{1}\stackrel{d}{\sim }$ gamma$\left( \theta \right) $. Let $\left( \xi
_{\left( k\right) },k\geq 1\right) $ be the ranked constitutive jumps of $%
\chi $ with $\chi \stackrel{d}{=}\sum_{k\geq 1}\xi _{\left( k\right) }$ and $%
\xi _{\left( 1\right) }>..>\xi _{\left( k\right) }>...$. Let $\left(
\varsigma _{\left( k\right) }:=\xi _{\left( k\right) }/\chi ;\text{ }k\geq
1\right) $ be the ranked normalized jumps, hence with $1=\sum_{k\geq
1}\varsigma _{\left( k\right) }$. Then, $\left( \varsigma _{\left( k\right)
};\text{ }k\geq 1\right) $ has Poisson-Dirichlet PD$\left( \theta \right) $
distribution, independent of $\chi =\chi _{1}$. This interpretation of $%
\left( \varsigma _{\left( k\right) };\text{ }k\geq 1\right) $ in terms of
normalized jumps of a gamma$\left( \theta \right) $-distributed random
variable is the limiting manifestation of the fact $\varsigma
_{k}=S_{k}/\left( S_{1}+..+S_{n}\right) $, $k=1,..,n$ (with $S_{k}$ iid
positive gamma$\left( \alpha \right) $-distributed random variables),
characterizing the Dirichlet D$_{n}\left( \alpha \right) $ model.\newline

\section{Partitioning Constructions Based on Integrable Infinitely Divisible
Random Variables with L\'{e}vy Measure Concentrated on $\left( 0,1\right) :$
the Unbounded Case}

\indent

Suppose some random resource (or energy, mass or amount of space) $\chi \geq
0$ is to be shared at random between (possibly infinitely many) species
(atoms, fragments), the amount of the individual share being necessarily
bounded by one. This random partitioning model can nicely be handled from
infinitely divisible (ID) random variables with L\'{e}vy measure
concentrated on $\left( 0,1\right) $ [See \cite{Ber1} and \cite{Steutel},
for general monographs on infinite-divisibility].

If the physical interpretation of $\chi $ is energy (as in earthquake
magnitude data with $\chi $ interpreting as the cumulative energy releases
on Earth over some period of time), our construction is, to some extent,
related to the Random Energy Model of Derrida (see \cite{Der} and \cite
{Derrida} and its ``Poissonian'' reformulation by \cite{Ruelle} and \cite
{Koukiou}). The partitioning nature of this problem is indeed well-known. In
insurance models, the individual share can represent the amount of a
particular claim resulting from some damage. In population genetics, it
could interpret as species abundance in a large population. In a partition
of mass problem, the share attributed to each of the constitutive element of
the partition is generally called the fragment size or mass. In an
economical context, the individuals share is their asset. In any case, the
peculiarity of our model is that the individual share of the constitutive
atoms of the partition are all physically necessarily bounded above.

\subsection{Random Partition of ``Energy'': The Model}

\indent

Let $\chi \geq 0$ be an infinitely divisible random variable with L\'{e}vy
measure for jumps $\Pi \left( dx\right) $ supported by $\left( 0,1\right) .$
Hence, with $\lambda \in {\Bbb R}$, let 
\begin{equation}
{\Bbb E}e^{-\lambda \chi }=\exp \left\{ -\int_{0}^{1}\left( 1-e^{-\lambda
x}\right) \Pi \left( dx\right) \right\}  \label{eq6}
\end{equation}
be the entire analytic Laplace-Stieltjes Transform (LST) of $\chi $s law. We
shall assume that ${\Bbb E}\chi =\theta <\infty $ so that $%
0<\int_{0}^{1}x\Pi \left( dx\right) =\theta <\infty $. We shall also assume
that $\Pi $ has a (continuous) density, say $\pi $. In this case, the
density $f_{\chi }$ of $\chi $ exists and is easily seen to solve the
functional equation 
\begin{equation}
xf_{\chi }\left( x\right) =\int_{0}^{x\wedge 1}f_{\chi }\left( x-z\right)
z\pi \left( z\right) dz.  \label{eq7}
\end{equation}
As is-well-known, the random variable $\chi $ is naturally associated to $%
\left( \chi _{t},t\geq 0\right) $ which is a process with stationary
independent increments. The process $\left( \chi _{t},t\geq 0\right) $ is a
subordinator with no drift and non-negative jumps restricted to $\left(
0,1\right) $ and $\chi =\chi _{1}$.\newline

Let $\overline{\Pi }\left( x\right) :=\int_{x}^{1}\Pi \left( dz\right) $.
Two cases arise, depending on whether $\overline{\Pi }\left( 0\right)
=\infty $ (the unbounded case) or $\overline{\Pi }\left( 0\right) <\infty $
(the bounded case). In this Section, we shall first address statistical
issues arising in the unbounded partitioning model.\newline

\begin{center}
{\bf Random Partition of Energy} $\chi $ {\bf (Unbounded Case): First
Properties}
\end{center}

Here $\overline{\Pi }\left( 0\right) =\infty $, where $\overline{\Pi }\left(
x\right) =\int_{x}^{1}\Pi \left( dz\right) $ is the tail of the L\'{e}vy
measure. In this case, the total mass of $\Pi $ is infinite and $\chi >0$.
Plainly, we have 
\begin{equation}
\chi \stackrel{d}{=}\sum_{k\geq 1}\overline{\Pi }^{-1}\left( S_{k}\right)
\label{eq8}
\end{equation}
where $\left( S_{k};k\geq 1\right) $ are points of a homogeneous Poisson
point process (PPP) on the half-line (with $T_{k}:=S_{k}-S_{k-1}$ iid and exp%
$\left( 1\right) $ distributed) and $\overline{\Pi }^{-1}$ the decreasing
inverse of $\overline{\Pi }.$

This decomposition constitutes a random partition of the random variable $%
\chi $ in terms of the (infinitely many) ranked constitutive jumps of $\chi
_{1}$, all bounded by $1$. Let $\xi _{\left( k\right) }:=\overline{\Pi }%
^{-1}\left( S_{k}\right) $, $k\geq 1$, be such $\left( 0,1\right) -$valued
jumps arranged in decreasing order $\xi _{\left( 1\right) }>..>\xi _{\left(
k\right) }>...$ They constitute a PPP on $\left( 0,1\right) $ with intensity 
$\Pi $, satisfying $\sum_{k\geq 1}{\Bbb E}\left( \xi _{\left( k\right)
}\right) =\theta $. In the decomposition of $\chi $ model, with $\chi 
\stackrel{d}{=}\sum_{k\geq 1}\xi _{\left( k\right) }$, the random variable $%
\xi _{\left( k\right) }$ interprets as the $k$th fragment size.

When $\theta =1$, the PPP system $\left( \xi _{\left( 1\right) },..,\xi
_{\left( k\right) },..\right) $ is said to be conservative in average. If $%
\theta >1$ ($\theta <1$) we shall say that the system is excessive
(defective).\newline

\begin{center}
{\bf First and Second Order Statistics: Correlation Structure}
\end{center}

We start with supplying easy informations.

First, from the law of large numbers $\frac{1}{k}\overline{\Pi }\left( \xi
_{\left( k\right) }\right) \rightarrow 1$ almost surely as $k\uparrow \infty 
$, supplying a useful information on the way $\xi _{\left( k\right) }$ goes
to $0$ when $k$ grows.

Next, the one-dimensional distribution of $\xi _{\left( k\right) }$ is
easily seen to be ${\Bbb P}\left( \xi _{\left( k\right) }\leq x\right) =%
{\Bbb P}\left( S_{k}>\overline{\Pi }\left( x\right) \right) =e^{-\overline{%
\Pi }\left( x\right) }\sum_{l=0}^{k-1}\frac{\overline{\Pi }\left( x\right)
^{l}}{l!}.$ In particular, 
\[
{\Bbb E}\left( \xi _{\left( k\right) }\right) =\frac{1}{\Gamma \left(
k\right) }\int_{0}^{\infty }\overline{\Pi }^{-1}\left( s\right)
s^{k-1}e^{-s}ds 
\]
\[
{\Bbb E}\left( \xi _{\left( k\right) }^{2}\right) =\frac{1}{\Gamma \left(
k\right) }\int_{0}^{\infty }\overline{\Pi }^{-1}\left( s\right)
^{2}s^{k-1}e^{-s}ds 
\]
\[
{\Bbb E}\left( \xi _{\left( k\right) }\xi _{\left( k+l\right) }\right)
=\int_{0}^{\infty }\int_{0}^{\infty }\overline{\Pi }^{-1}\left( s_{1}\right) 
\overline{\Pi }^{-1}\left( s_{1}+s_{2}\right) \frac{%
s_{1}^{k-1}e^{-s_{1}}s_{2}^{l-1}e^{-s_{2}}}{\Gamma \left( k\right) \Gamma
\left( l\right) }ds_{1}ds_{2} 
\]
are the first, second moments of $\xi _{\left( k\right) }$, together with
the joint second moment of $\xi _{\left( k\right) }$ and $\xi _{\left(
k+l\right) }$, $k$ and $l\geq 1$. From this last expression, there is no
general stationarity property to be expected. As a result, with $w_{k}:=%
{\Bbb E}\xi _{\left( k\right) }$, the second order quantity and its $\left(
q_{1},q_{2}\right) -$definition domain 
\begin{equation}
C_{l}\left( q_{1},q_{2}\right) :={\Bbb E}\left( \sum_{k\geq 1}w_{k}\xi
_{\left( k\right) }^{q_{1}}\xi _{\left( k+l\right) }^{q_{2}}\right)
\label{e1}
\end{equation}
deserves some interest. It gives weight $w_{k}$ to the $k$th contribution to
the full pair-correlation function at distance $l$.\newline

{\em Example }(Dickman){\bf :}

Assume $\Pi \left( dx\right) =\frac{\theta }{x}{\bf I}_{x\in \left(
0,1\right) }dx$, then $\overline{\Pi }^{-1}\left( s\right) =\exp \left\{
-s/\theta \right\} $. Then $\chi $s distribution is closely related to
Dickman distribution (see below). In this case, 
\begin{eqnarray*}
{\Bbb E}\left( \xi _{\left( k\right) }\right) &=&\left( \frac{\theta }{%
\theta +1}\right) ^{k}\text{, }\sigma ^{2}\left( \xi _{\left( k\right)
}\right) =\left( \frac{\theta }{\theta +2}\right) ^{k}-\left( \frac{\theta }{%
\theta +1}\right) ^{2k}, \\
{\Bbb E}\left( \xi _{\left( k\right) }\xi _{\left( k+l\right) }\right)
&=&\left( \frac{\theta }{\theta +2}\right) ^{k}\left( \frac{\theta }{\theta
+1}\right) ^{l}
\end{eqnarray*}
are the mean, variance of $\xi _{\left( k\right) }$ and correlation of $\xi
_{\left( k\right) },$ $\xi _{\left( k+l\right) }$. In this particular
separable case, the covariance coefficient $cov\left( \xi _{\left( k\right)
},\xi _{\left( k+l\right) }\right) =\sigma ^{2}\left( \xi _{\left( k\right)
}\right) \left( \frac{\theta }{\theta +1}\right) ^{l}$ and 
\[
{\Bbb E}\left( \sum_{k\geq 1}\xi _{\left( k\right) }\xi _{\left( k+l\right)
}\right) =\frac{\theta }{2}\left( \frac{\theta }{\theta +1}\right) ^{l} 
\]
has exponential decay with $l$. More generally, in this particular case,
with definition domain $q_{1}+q_{2}>-\theta /\left( \theta +1\right) $ and $%
q_{2}>-\theta $, we obtain 
\begin{equation}
C_{l}\left( q_{1},q_{2}\right) =\frac{\theta ^{2}}{\theta +\left( \theta
+1\right) \left( q_{1}+q_{2}\right) }\left( \frac{\theta }{\theta +q_{2}}%
\right) ^{l}.  \label{e2}
\end{equation}
This particular Dickman-model exhibits many other remarkable properties;
these will be emphasized in some detail in the sequel. $\square $\newline

\begin{center}
{\bf Campbell Formula}
\end{center}

A very useful formula in our context is Campbell formula. We first recall it
and then show its usefulness in the computation of statistical variables of
concrete interest in the partitioning problem under study.

Let $\lambda \geq 0$ and $g$ be a measurable function such that $%
\int_{0}^{1}\left( 1-e^{-\lambda g\left( x\right) }\right) \Pi \left(
dx\right) <\infty $, then by Campbell formula (see \cite{Neveu}) 
\begin{equation}
{\Bbb E}\exp \left\{ -\lambda \sum_{k\geq 1}g\left( \overline{\Pi }%
^{-1}\left( S_{k}\right) \right) \right\} =\exp \left\{ -\int_{0}^{1}\left(
1-e^{-\lambda g\left( x\right) }\right) \Pi \left( dx\right) \right\}
\label{eq9}
\end{equation}
is the Laplace-Stieltjes transform (LST) of $\sum_{k\geq 1}g\left( \overline{%
\Pi }^{-1}\left( S_{k}\right) \right) $. In particular, its mean value is 
\[
{\Bbb E}\sum_{k\geq 1}g\left( \overline{\Pi }^{-1}\left( S_{k}\right)
\right) =\int_{0}^{1}g\left( x\right) \Pi \left( dx\right) . 
\]
Let us draw some conclusions of these elementary facts.\newline

\begin{center}
{\bf Cumulative energy of the }$K-${\bf biggest and of the remaining events}
\end{center}

Let $K\geq 1$. The random variable $\chi _{K}^{-}:=\sum_{k>K}\xi _{\left(
k\right) }$ represents the amount of total energy $\chi $ concentrated in
the lowest energy levels (at rank $K+1$ and below). Let us consider the
problem of computing its law. We clearly have $\chi _{K}^{-}\stackrel{d}{=}%
\sum_{k\geq 1}\overline{\Pi }^{-1}\left( S_{K}+S_{k}\right) $ where the PPP $%
\left( S_{k};k\geq 1\right) $ is independent of $S_{K}\stackrel{d}{\sim }$
gamma$\left( K\right) $. As a result, applying Campbell formula (\ref{eq9})
with $g\left( x\right) =\overline{\Pi }^{-1}\left( S_{K}+\overline{\Pi }%
\left( x\right) \right) $, we obtain 
\[
{\Bbb E}\exp \left\{ -\lambda \chi _{K}^{-}\right\} ={\Bbb E}\exp \left\{
-\lambda \sum_{k\geq 1}\overline{\Pi }^{-1}\left( S_{K}+\overline{\Pi }%
\left( \xi _{\left( k\right) }\right) \right) \right\} 
\]
\begin{equation}
={\Bbb E}\exp \left\{ -\int_{0}^{1}\left( 1-e^{-\lambda \overline{\Pi }%
^{-1}\left( S_{K}+\overline{\Pi }\left( x\right) \right) }\right) \Pi \left(
dx\right) \right\} ,  \label{e2a}
\end{equation}
where the last expectation is over $S_{K}$. Note that $\chi =\chi
_{K}^{+}+\chi _{K}^{-}$ where $\chi _{K}^{+}:=\sum_{k=1}^{K}\xi _{\left(
k\right) }$ is the contribution of the $K$ largest energy levels to $\chi .$

This question is closely related to the following problem: let $x>0$ be some
threshold value. Define 
\[
K\left( x\right) :=\inf \left( K\geq 1:\chi _{K}^{+}>x\right) 
\]
to be the first time the cumulated fragments size of $\left( \xi _{\left(
k\right) },k\geq 1\right) $ exceeds $x.$ Then, ${\Bbb P}\left( K\left(
x\right) >K\right) ={\Bbb P}\left( \chi _{K}^{+}\leq x\right) $ and the
distribution of $K\left( x\right) $ results from the one of $\chi _{K}^{+}.$
See \cite{Lindenberg} for similar considerations.\newline

{\em Example }(Dickman){\em :}

Assuming $\Pi \left( dx\right) =\frac{\theta }{x}{\bf I}_{x\in \left(
0,1\right) }dx$, then $\overline{\Pi }^{-1}\left( s\right) =\exp \left\{
-s/\theta \right\} $ and 
\begin{eqnarray*}
{\Bbb E}\exp \left\{ -\lambda \chi _{K}^{-}\right\} &=&{\Bbb E}\exp \left\{
-\int_{0}^{1}\left( 1-e^{-\lambda \overline{\Pi }^{-1}\left( S_{K}+\overline{%
\Pi }\left( x\right) \right) }\right) \Pi \left( dx\right) \right\} \\
&=&{\Bbb E}\exp \left\{ -\theta \int_{0}^{1}\left( 1-e^{-\lambda xe^{-\frac{%
S_{K}}{\theta }}}\right) \frac{1}{x}dx\right\} .
\end{eqnarray*}
This shows that, in this particular case, 
\[
\chi _{K}^{-}\stackrel{d}{=}R_{K}\cdot \chi \text{ and }\chi _{K}^{+}%
\stackrel{d}{=}\left( 1-R_{K}\right) \cdot \chi 
\]
where $R_{K}:=\exp \left\{ -\frac{S_{K}}{\theta }\right\} \in \left(
0,1\right) $ is log-gamma$\left( K,\theta \right) $ distributed, with ${\Bbb %
E}R_{K}^{q}=\left[ \theta /\left( \theta +q\right) \right] ^{K}$,
independent of $\chi .$ When $K=\left[ \left( \log _{2}\left( 1+1/\theta
\right) \right) ^{-1}\right] $, the average wealth ${\Bbb E}\chi _{K}^{-}$
is half the one of $\chi $.

In this example, we shall show below that $\chi $ has a Dickman type
distribution, see (\ref{eq28}) below, resulting in an intricate distribution
for $\chi _{K}^{-}$ and $\chi _{K}^{+}$ and consequently of $K\left(
x\right) $. $\square $\newline

\begin{center}
{\bf Partition Function of} $\chi $
\end{center}

Partition functions of energy are interesting quantities. Taking in
particular $g\left( x\right) =x^{\beta }$ in (\ref{eq9}), the full
Laplace-Stieltjes transform of $\sum_{k\geq 1}\xi _{\left( k\right) }^{\beta
}$ is obtained as 
\begin{equation}
{\Bbb E}\exp \left\{ -\lambda \sum_{k\geq 1}\xi _{\left( k\right) }^{\beta
}\right\} =\exp \left\{ -\int_{0}^{1}\left( 1-e^{-\lambda x^{\beta }}\right)
\Pi \left( dx\right) \right\} .  \label{eq10}
\end{equation}
Its mean value 
\begin{equation}
\phi \left( \beta \right) :={\Bbb E}\sum_{k\geq 1}\xi _{\left( k\right)
}^{\beta }=\int_{0}^{1}x^{\beta }\Pi \left( dx\right)  \label{eq11}
\end{equation}
is defined for values of $\beta >\beta _{*}$ for which $\int_{0}^{1}x^{\beta
}\Pi \left( dx\right) <\infty $, with 
\[
\beta _{*}:=\sup \left( \beta :\phi \left( \beta \right) =\infty \right) \in
\left[ 0,1\right) . 
\]
Note indeed that $\phi \left( 0\right) =\infty $ and $\phi \left( 1\right)
=\theta .$\newline

In this setup, the L\'{e}vy measure $\Pi $ interprets as follows.

Let $N_{+}\left( x\right) :=\sum_{k\geq 1}{\bf I}\left( \xi _{\left(
k\right) }>x\right) $ be the random number of $\xi _{\left( k\right) }$
exceeding $x\in \left( 0,1\right) $. Then $\overline{\Pi }\left( x\right) $
is the expected value of this number. Indeed 
\begin{eqnarray*}
\overline{\Pi }\left( x\right) &=&\sum_{k\geq 1}{\Bbb P}\left( \xi _{\left(
k\right) }>x\right) =\sum_{k\geq 1}{\Bbb P}\left( S_{k}\leq \overline{\Pi }%
\left( x\right) \right) \\
&=&e^{-\overline{\Pi }\left( x\right) }\sum_{k\geq 1}\sum_{l\geq k}\frac{%
\overline{\Pi }\left( x\right) ^{l}}{l!}=e^{-\overline{\Pi }\left( x\right)
}\sum_{l\geq 1}\frac{\overline{\Pi }\left( x\right) ^{l}}{\left( l-1\right) !%
}.
\end{eqnarray*}
The random variable $\sum_{k\geq 1}\xi _{\left( k\right) }^{\beta }$ is
called the partition function of $\chi $ and $\Pi $ its structural (or
occupation) measure.\newline

\begin{center}
{\bf The Numbers of Atoms of} $\chi $ {\bf above Cutoff }$\epsilon $ {\bf %
and the Contribution to Total Mass of those Atoms above and below} $\epsilon 
${\bf : Threshold Statistics}
\end{center}

The above considerations naturally suggest the following problems of
interest in Statistics.

\begin{itemize}
\item  Upper-threshold Statistics.
\end{itemize}

If $\epsilon \in \left( 0,1\right) $ is some cutoff or threshold value, let $%
N_{+}\left( \epsilon \right) $ count the numbers of atoms of the partition
of $\chi $ exceeding $\epsilon $. If $\chi $ is the amount of some natural
resource to be shared between infinitely many agents on the market, $%
\epsilon $ stands for the minimal individual wealth below which each agent
should be considered as indigent (e.g. below the poverty line). If $\chi $
stands for ``energy'', $\epsilon $ could interpret as the level below which
micro-events are undetectable by the currently available measuring devices
(if one thinks of a sequence of earthquakes magnitude data for example).%
\newline

By Campbell formula 
\[
{\Bbb E}\exp \left\{ -\lambda N_{+}\left( \epsilon \right) \right\} =\exp
\left\{ -\int_{0}^{1}\left( 1-e^{-\lambda {\bf I}\left( x>\epsilon \right)
}\right) \Pi \left( dx\right) \right\} 
\]
\begin{equation}
=\exp \left\{ -\overline{\Pi }\left( \epsilon \right) \left( 1-e^{-\lambda
}\right) \right\}  \label{eq12}
\end{equation}
is the full Laplace-Stieltjes transform of $N_{+}\left( \epsilon \right) $.
This shows that $N_{+}\left( \epsilon \right) $ is in fact Poisson
distributed with intensity $\overline{\Pi }\left( \epsilon \right) $.
Recalling $\overline{\Pi }\left( \epsilon \right) \stackunder{\epsilon
\downarrow 0}{\rightarrow }\infty $, the law of large numbers gives 
\begin{equation}
N_{+}\left( \epsilon \right) /\overline{\Pi }\left( \epsilon \right) 
\stackrel{a.s.}{\rightarrow }1\text{, }\epsilon \downarrow 0.  \label{eq13}
\end{equation}
That $N_{+}\left( \epsilon \right) $ is Poisson distributed may be also
checked as follows: we have $N_{+}\left( \epsilon \right) =\inf \left( k\geq
1:\xi _{\left( k\right) }\leq \epsilon \right) -1$ and ${\Bbb P}\left(
N_{+}\left( \epsilon \right) \geq k\right) ={\Bbb P}\left( \xi _{\left(
k\right) }>\epsilon \right) $. This is also ${\Bbb P}\left( S_{k}\leq 
\overline{\Pi }\left( \epsilon \right) \right) =e^{-\overline{\Pi }\left(
\epsilon \right) }\sum_{l\geq k}\frac{\overline{\Pi }\left( \epsilon \right)
^{l}}{l!}$ and $N_{+}\left( \epsilon \right) $ is Poisson with intensity $%
\overline{\Pi }\left( \epsilon \right) $.\newline

{\em Remark} (randomization of the cutoff):

A slightly more general problem is to consider the random variable $%
N_{+}\left( \epsilon _{1}\right) :=\sum_{k\geq 1}{\bf I}\left( \xi _{\left(
k\right) }>\epsilon _{k}\right) $ where $\left( \epsilon _{k}\text{, }k\geq
1\right) $ are iid $\left( 0,1\right) -$valued random variables, independent
of $\left( \xi _{\left( k\right) }\text{, }k\geq 1\right) $. In this model,
the poverty threshold attached to each agent is assumed random but drawn
from the same distribution and with mutual independence.

From Campbell formula, we obtain 
\[
{\Bbb E}\exp \left\{ -\lambda N_{+}\left( \epsilon _{1}\right) \right\}
=\exp \left\{ -{\Bbb E}\overline{\Pi }\left( \epsilon _{1}\right) \left(
1-e^{-\lambda }\right) \right\} , 
\]
showing that $N_{+}\left( \epsilon _{1}\right) $ is Poisson distributed with
intensity ${\Bbb E}\overline{\Pi }\left( \epsilon _{1}\right) $ if ${\Bbb E}%
\overline{\Pi }\left( \epsilon _{1}\right) <\infty .$

This is useful in the problem of random covering of $\left( \xi _{\left(
k\right) }\text{, }k\geq 1\right) $ by random intervals with sizes $\left(
\epsilon _{k}\text{, }k\geq 1\right) $. In particular, the covering
probability is 
\[
{\Bbb P}\left( N_{+}\left( \epsilon _{1}\right) =0\right) =\exp \left\{ -%
{\Bbb E}\overline{\Pi }\left( \epsilon _{1}\right) \right\} . 
\]
Assuming (Dickman): $\overline{\Pi }\left( \epsilon \right) =-\theta \log
\epsilon $ and $\epsilon _{1}\stackrel{d}{\sim }$ Uniform$\left( 0,1\right) $%
, the intensity reads ${\Bbb E}\overline{\Pi }\left( \epsilon _{1}\right)
=-\theta \int_{0}^{1}\log \epsilon d\epsilon =\theta $. In this case, $%
N_{+}\left( \epsilon _{1}\right) $ simply is Poisson$\left( \theta \right) $
distributed. $\square $\newline

The contribution to total mass $\chi $ of those $\xi _{\left( k\right) }$
above $\epsilon \in \left( 0,1\right) $, which is 
\[
\chi _{+}\left( \epsilon \right) :=\sum_{k\geq 1}\xi _{\left( k\right) }{\bf %
I}\left( \xi _{\left( k\right) }>\epsilon \right) =\sum_{k=1}^{N_{+}\left(
\epsilon \right) }\xi _{\left( k\right) }, 
\]
is such that 
\begin{eqnarray*}
{\Bbb E}\exp \left\{ -\lambda \chi _{+}\left( \epsilon \right) \right\}
&=&\exp \left\{ -\int_{0}^{1}\left( 1-e^{-\lambda x{\bf I}\left( x>\epsilon
\right) }\right) \Pi \left( dx\right) \right\} \\
&=&\exp \left\{ -\int_{\epsilon }^{1}\left( 1-e^{-\lambda x}\right) \Pi
\left( dx\right) \right\} .
\end{eqnarray*}
This is the LST of an infinitely divisible (ID) random variable with
L\'{e}vy measure $\Pi $ concentrated on $\left( \epsilon ,1\right) $ for
which clearly $\chi _{+}\left( \epsilon \right) \stackrel{d}{\rightarrow }%
\chi $ ($\epsilon \downarrow 0$). This random variable is of the compound
Poisson type since 
\begin{equation}
{\Bbb E}\exp \left\{ -\lambda \chi _{+}\left( \epsilon \right) \right\}
=\exp \left\{ -\overline{\Pi }\left( \epsilon \right) \left(
1-\int_{\epsilon }^{1}e^{-\lambda x}\Pi \left( dx\right) /\overline{\Pi }%
\left( \epsilon \right) \right) \right\} .  \label{eq14}
\end{equation}
Here indeed, $F_{\epsilon }\left( dx\right) :=\Pi \left( dx\right) /%
\overline{\Pi }\left( \epsilon \right) $ is a probability distribution.
Hence, with $u_{k},$ $k\geq 1$ an iid $\left( 0,1\right) -$valued uniform
sequence, $P_{\epsilon }$ a Poisson random variable with intensity $%
\overline{\Pi }\left( \epsilon \right) $, $\chi _{+}\left( \epsilon \right)
=\sum_{k=1}^{P_{\epsilon }}\overline{F}_{\epsilon }^{-1}\left( u_{k}\right) $
belongs to the class of compound Poisson random variables. Stated
differently 
\begin{equation}
\chi _{+}\left( \epsilon \right) =\sum_{k=1}^{P_{\epsilon }}\overline{F}%
_{\epsilon }^{-1}\left( u_{\left( k\right) ,P_{\epsilon }}\right)
\label{eq15}
\end{equation}
where $u_{\left( 1\right) ,P_{\epsilon }}<..<u_{\left( P_{\epsilon }\right)
,P_{\epsilon }}$ is obtained from uniform sample $u_{1},..,u_{P_{\epsilon }}$
while ordering the constitutive terms. We note that $\chi _{+}\left(
\epsilon \right) $ has an atom at $\chi _{+}\left( \epsilon \right) =0$ with
probability $e^{-\overline{\Pi }\left( \epsilon \right) }$. This partition
is also $\chi _{+}\left( \epsilon \right) \stackrel{d}{=}\sum_{k=1}^{P_{%
\epsilon }}\overline{\Pi }^{-1}\left( \overline{\Pi }\left( \epsilon \right)
u_{\left( k\right) ,P_{\epsilon }}\right) $ where, when $\epsilon \downarrow
0$, $P_{\epsilon }\stackrel{a.s.}{\rightarrow }\infty $ and $\left( 
\overline{\Pi }\left( \epsilon \right) u_{\left( 1\right) ,P_{\epsilon }},..,%
\overline{\Pi }\left( \epsilon \right) u_{\left( k\right) ,P_{\epsilon
}}\right) \stackrel{d}{\rightarrow }\left( S_{1},..,S_{k},..\right) $ a
Poisson point process on ${\Bbb R}^{+}.$ Thus the decomposition of $\chi
_{+}\left( \epsilon \right) $ constitutes a weak Poisson-partition
approximation to the one of $\chi $.\newline

{\em Remarks:}

$\left( i\right) $ A slightly more general problem is to consider the random
variable $\chi _{+}\left( \epsilon _{1}\right) :=\sum_{k\geq 1}\xi _{\left(
k\right) }{\bf I}\left( \xi _{\left( k\right) }>\epsilon _{k}\right) $ where 
$\left( \epsilon _{k}\text{, }k\geq 1\right) $ are iid $\left( 0,1\right) -$%
valued random variables, independent of $\left( \xi _{\left( k\right) }\text{%
, }k\geq 1\right) $. From Campbell formula, performing an integration by
parts, with $F_{\epsilon _{1}}\left( \epsilon \right) ={\Bbb P}\left(
\epsilon _{1}\leq \epsilon \right) $, we obtain 
\[
{\Bbb E}\exp \left\{ -\lambda \chi _{+}\left( \epsilon _{1}\right) \right\}
=\exp \left\{ -{\Bbb E}\int_{\epsilon _{1}}^{1}\left( 1-e^{-\lambda
x}\right) \Pi \left( dx\right) \right\} 
\]
\begin{equation}
=\exp \left\{ -\int_{0}^{1}\left( 1-e^{-\lambda x}\right) F_{\epsilon
_{1}}\left( x\right) \Pi \left( dx\right) \right\}  \label{e3}
\end{equation}
showing that, in general, $\chi _{+}\left( \epsilon _{1}\right) $ is an ID
random variable with L\'{e}vy measure for jumps $F_{\epsilon _{1}}\left(
x\right) \Pi \left( dx\right) .$

Note also that if $\chi _{-}\left( \epsilon _{1}\right) :=\sum_{k\geq 1}\xi
_{\left( k\right) }{\bf I}\left( \xi _{\left( k\right) }\leq \epsilon
_{k}\right) $, clearly 
\begin{equation}
{\Bbb E}\exp \left\{ -\lambda \chi _{-}\left( \epsilon _{1}\right) \right\}
=\exp \left\{ -\int_{0}^{1}\left( 1-e^{-\lambda x}\right) \overline{F}%
_{\epsilon _{1}}\left( x\right) \Pi \left( dx\right) \right\}  \label{e4}
\end{equation}
where $\overline{F}_{\epsilon _{1}}\left( x\right) :=1-F_{\epsilon
_{1}}\left( x\right) .$\newline

$\left( ii\right) $ Finally, in the random covering of $\left( \xi _{\left(
k\right) }\text{, }k\geq 1\right) $ by random intervals $\left( \epsilon _{k}%
\text{, }k\geq 1\right) $ context, the quantity $\chi _{g}:=\sum_{k\geq
1}\left( \xi _{\left( k\right) }-\epsilon _{k}\right) _{+}$ interprets as
the total gaps' length (in the economical context, it is the excess-wealth
of the well-off agents). We obtain directly 
\[
{\Bbb E}\exp \left\{ -\lambda \sum_{k\geq 1}\left( \xi _{\left( k\right)
}-\epsilon _{k}\right) _{+}\right\} =\exp \left\{ -{\Bbb E}\int_{\epsilon
_{1}}^{1}\left( 1-e^{-\lambda \left( x-\epsilon _{1}\right) }\right) \Pi
\left( dx\right) \right\} 
\]
\begin{equation}
=\exp \left\{ -\int_{0}^{1}\left( 1-e^{-\lambda z}\right) {\Bbb E}\Pi
_{\epsilon _{1}}\left( dz\right) \right\} ,  \label{e5}
\end{equation}
where $\Pi _{\epsilon _{1}}\left( dz\right) $ is the image measure of $\Pi
\left( dx\right) $ by the application $x\rightarrow z=x-\epsilon _{1}\in
\left( 0,1-\epsilon _{1}\right) $. This is the LST of an ID random variable
with L\'{e}vy measure for jumps ${\Bbb E}\Pi _{\epsilon _{1}}\left(
dz\right) $. $\square $\newline

{\em Examples }(Dickman){\em :}

Assuming $\Pi \left( dx\right) =\frac{\theta }{x}{\bf I}_{x\in \left(
0,1\right) }dx,$ we have $\Pi _{\epsilon _{1}}\left( dz\right) =\frac{\theta 
}{z+\epsilon _{1}}{\bf I}_{z\in \left( 0,1-\epsilon _{1}\right) }dz.$

$\bullet $ If in addition, $\epsilon _{1}$ is uniformly distributed on $%
\left( 0,1\right) $, we find explicitly 
\[
{\Bbb E}\Pi _{\epsilon _{1}}\left( dz\right) =dz\int_{0}^{1-z}\frac{\theta }{%
z+\epsilon }d\epsilon =-\theta \log zdz,z\in \left( 0,1\right) . 
\]
In the chosen example, the total gaps' length is an ID (rate $\theta $
compound Poisson) random variable with logarithmic density for jumps $-\log z%
{\bf I}_{z\in \left( 0,1\right) }$ and ${\Bbb E}\chi _{g}/{\Bbb E}\chi
=-\int_{0}^{1}x\log xdx<1$ is the average reduction factor$.$

$\bullet $ If $\epsilon _{1}$ is not random, with $\epsilon _{1}\stackrel{d}{%
\sim }$ $\delta _{\epsilon _{1}-\epsilon }$, the total gaps' length is a
(rate $-\theta \log \epsilon $) compound Poisson random variable with
density for jumps: $\frac{-1}{\left( z+\epsilon \right) \log \epsilon }{\bf I%
}_{z\in \left( 0,1-\epsilon \right) }.$ In addition, one gets ${\Bbb E}\chi
_{g}/{\Bbb E}\chi =1-\epsilon +\epsilon \log \epsilon \rightarrow 1$ ($%
\epsilon \downarrow 0^{+}$).\newline

Incidentally, note that the lack of wealth of the poorest, which is $%
\sum_{k\geq 1}\left( \epsilon _{k}-\xi _{\left( k\right) }\right) _{+}$,
diverges. $\square $\newline

\begin{itemize}
\item  Sub-threshold Statistics.
\end{itemize}

Similarly, let $N_{-}\left( \epsilon \right) :=\sum_{k\geq 1}{\bf I}\left(
\xi _{\left( k\right) }\leq \epsilon \right) $ count the random number of $%
\xi _{\left( k\right) }$ below cutoff $\epsilon \in \left( 0,1\right) ,$
then $N_{-}\left( \epsilon \right) =\infty $ for all such $\epsilon .$ The
contribution to total mass $\chi $ of those $\xi _{\left( k\right) }$ below $%
\epsilon \in \left( 0,1\right) $, which is 
\[
\chi _{-}\left( \epsilon \right) :=\sum_{k\geq 1}\xi _{\left( k\right) }{\bf %
I}\left( \xi _{\left( k\right) }\leq \epsilon \right) =\sum_{k>N_{+}\left(
\epsilon \right) }\xi _{\left( k\right) } 
\]
is such that 
\begin{eqnarray*}
{\Bbb E}\exp \left\{ -\lambda \chi _{-}\left( \epsilon \right) \right\}
&=&\exp \left\{ -\int_{0}^{1}\left( 1-e^{-\lambda x{\bf I}\left( x\leq
\epsilon \right) }\right) \Pi \left( dx\right) \right\} \\
&=&\exp \left\{ -\int_{0}^{\epsilon }\left( 1-e^{-\lambda x}\right) \Pi
\left( dx\right) \right\} .
\end{eqnarray*}
This is the LST of an ID random variable with L\'{e}vy measure $\Pi $
concentrated on $\left( 0,\epsilon \right) $, showing that $\chi _{-}\left(
\epsilon \right) $ and $\chi _{+}\left( \epsilon \right) $ are independent
with $\chi \stackrel{d}{=}\chi _{-}\left( \epsilon \right) +$ $\chi
_{+}\left( \epsilon \right) $. Furthermore, ${\Bbb E}\chi _{-}\left(
\epsilon \right) =\int_{0}^{\epsilon }x\Pi \left( dx\right) \sim \epsilon
^{2}\pi \left( \epsilon \right) \rightarrow _{\epsilon \downarrow 0}0$ and
the variance $\sigma ^{2}\left[ \chi _{-}\left( \epsilon \right) \right]
\sim \epsilon ^{3}\pi \left( \epsilon \right) \rightarrow _{\epsilon
\downarrow 0}0$. As a result,\newline

$*$ If $\sigma \left[ \chi _{-}\left( \epsilon \right) \right] /{\Bbb E}\chi
_{-}\left( \epsilon \right) \stackunder{\epsilon \downarrow 0}{\rightarrow }%
0 $, one can check that the Central Limit Theorem holds 
\begin{equation}
\frac{\chi _{-}\left( \epsilon \right) -{\Bbb E}\chi _{-}\left( \epsilon
\right) }{\sigma \left[ \chi _{-}\left( \epsilon \right) \right] }\stackrel{d%
}{\stackunder{\epsilon \downarrow 0}{\rightarrow }}{\cal N}\left( 0,1\right)
.  \label{eq16}
\end{equation}

$*$ If $\epsilon \pi \left( \epsilon \right) \rightarrow _{\epsilon
\downarrow 0}a>0$, then $\sigma \left[ \chi _{-}\left( \epsilon \right)
\right] /{\Bbb E}\chi _{-}\left( \epsilon \right) \stackunder{\epsilon
\downarrow 0}{\rightarrow }1/\sqrt{a}$ and we easily find 
\begin{equation}
\frac{\chi _{-}\left( \epsilon \right) }{{\Bbb E}\chi _{-}\left( \epsilon
\right) }\stackrel{d}{\stackunder{\epsilon \downarrow 0}{\rightarrow }}\chi
_{a}\text{ with }{\Bbb E}e^{-\lambda \chi _{a}}=e^{-a\int_{0}^{1}\left(
1-e^{-\lambda x/a}\right) dx/x}.  \label{eq17}
\end{equation}
The limiting random variable $\chi _{a}$ is (mean $1$) infinitely divisible
with major interest. It will be studied in some detail in the sequel (see
Subsection $4.1$).

Although the atoms below the cutoff are infinitely many, their contribution
to total mass always goes to $0$ as the cutoff approaches $0$.\newline

\begin{center}
{\bf Weighted Partitions (Modulation)}
\end{center}

Let $\left( \mu _{k};k\geq 1\right) $ be a sequence of iid non-negative
random variables, independent of $\left( \xi _{\left( k\right) };k\geq
1\right) $. We shall investigate some properties of the weighted or
modulated random sequence $\left( \mu _{k}\xi _{\left( k\right) };k\geq
1\right) $ as a new random transformed partition of $\chi _{\mu
_{1}}:=\sum_{k\geq 1}\mu _{k}\xi _{\left( k\right) }.$ This question appears
in the following problem: assume the events $\left( \xi _{\left( k\right)
};k\geq 1\right) $, summing up to $\chi $, are each corrupted by some
multiplicative independent noise $\left( \mu _{k};k\geq 1\right) $; then the
observed sequence of events becomes $\left( \mu _{k}\xi _{\left( k\right)
};k\geq 1\right) $ and the observed cumulative energy turns out to be $\chi
_{\mu _{1}}$.

Let us first consider the quantity ${\Bbb E}\exp \left\{ -\lambda
\sum_{k\geq 1}g\left( \mu _{k}\xi _{\left( k\right) }\right) \right\} $.
From Campbell formula, we have 
\[
{\Bbb E}\exp \left\{ -\lambda \sum_{k\geq 1}g\left( \mu _{k}\xi _{\left(
k\right) }\right) \right\} ={\Bbb E}\left\{ \prod_{k\geq 1}{\Bbb E}\left(
\exp -\lambda g\left( \mu _{k}\xi _{\left( k\right) }\right) \mid \xi
_{\left( k\right) }\right) \right\} 
\]
\[
=\exp \left\{ -{\Bbb E}\int_{0}^{1}\left( 1-e^{-\lambda g\left( \mu
_{1}x\right) }\right) \Pi \left( dx\right) \right\} 
\]
\begin{equation}
=\exp \left\{ -{\Bbb E}\int_{0}^{\mu _{1}}\left( 1-e^{-\lambda g\left(
z\right) }\right) \Pi _{\mu _{1}}\left( dz\right) \right\}  \label{e6}
\end{equation}
with $\Pi _{\mu _{1}}\left( dz\right) $ the image measure of $\Pi \left(
dx\right) $ by the application $x\rightarrow z=\mu _{1}x$.\newline
{\em Examples }(Dickman){\em :}

We note the scale-invariance property $\Pi _{\mu _{1}}\left( dz\right) =\Pi
\left( dz\right) $ when $\Pi \left( dx\right) =\frac{\theta }{x}dx$. Using
an integration by parts and putting $\overline{F}_{\mu _{1}}\left( x\right) =%
{\Bbb P}\left( \mu _{1}>x\right) $, this shows that in this particular case
for $\Pi $ only 
\begin{equation}
{\Bbb E}\exp \left\{ -\lambda \sum_{k\geq 1}g\left( \mu _{k}\xi _{\left(
k\right) }\right) \right\} =e^{-\int_{0}^{\infty }\left( 1-e^{-\lambda
g\left( x\right) }\right) \overline{F}_{\mu _{1}}\left( x\right) \Pi \left(
dx\right) }.  \label{e7}
\end{equation}
In particular, $\sum_{k\geq 1}\mu _{k}\xi _{\left( k\right) }$ is a positive
ID random variable with no negative jumps whose induced L\'{e}vy measure for
jumps is $\frac{\theta \overline{F}_{\mu _{1}}\left( x\right) }{x}dx.$ For
example

$\left( i\right) $ $p$-thinning: if $\mu _{1}\stackrel{d}{\sim }$ Bernoulli$%
\left( p\right) $, the new L\'{e}vy measure is $\frac{p\theta }{x}{\bf I}%
_{x\in \left( 0,1\right) }dx.$

$\left( ii\right) $ uniform thinning: if $\mu _{1}\stackrel{d}{\sim }$
Uniform$\left( 0,1\right) $, the new transformed L\'{e}vy measure is $\frac{%
\theta \left( 1-x\right) }{x}{\bf I}_{x\in \left( 0,1\right) }dx.$

$\left( iii\right) $ exponential scaling: if $\mu _{1}\stackrel{d}{\sim }$
exp$\left( 1\right) $, the new L\'{e}vy measure is $\frac{\theta }{x}e^{-x}%
{\bf I}_{x>0}dx$ which is the one of a gamma (or Moran) subordinator. Note
that $\mu _{1}\xi _{\left( 1\right) }\succeq _{st}..\succeq _{st}\mu _{k}\xi
_{\left( k\right) }\succeq _{st}..$and that, although the sequence $\left(
\mu _{k}\xi _{\left( k\right) };k\geq 1\right) $ is not strongly ordered by
decreasing sizes, the constitutive terms sum up to a gamma$\left( \theta
\right) -$distributed random variable. This should not be confused with the
other natural partition of $\chi \stackrel{d}{\sim }$ gamma$\left( \theta
\right) $ given by 
\[
\chi \stackrel{d}{=}\sum_{k\geq 1}\varsigma _{\left( k\right) }\text{, with }%
\varsigma _{\left( 1\right) }>..>\varsigma _{\left( k\right) }>.. 
\]
where $\varsigma _{\left( k\right) }=\overline{\Pi }^{-1}\left( S_{k}\right) 
$, $k\geq 1$ and $\overline{\Pi }\left( x\right) =\int_{x}^{\infty }\frac{%
\theta }{z}\exp \left\{ -z\right\} dz$. This constitutes an example where
two distinct sequences $\left( \mu _{k}\xi _{\left( k\right) };k\geq
1\right) $ and $\left( \varsigma _{\left( k\right) };k\geq 1\right) $ both
share the same partition function. $\square $\newline

\begin{center}
{\bf Typical Fragment Size from} $\left( \xi _{\left( k\right) },k\geq
1\right) $
\end{center}

We can define the typical fragment size from $\left( \xi _{\left( k\right)
},k\geq 1\right) $ to be a $\left( 0,1\right) -$valued random variable, say $%
\xi $, with density $f_{\xi }\left( x\right) $, whose distribution function $%
F_{\xi }\left( x\right) $ is defined by the random mixture 
\[
F_{\xi }\left( s\right) =\sum_{k\geq 1}w_{k}F_{\xi _{\left( k\right)
}}\left( s\right) . 
\]
Here, weights $w_{k}=\frac{1}{\theta }{\Bbb E}\left( \xi _{\left( k\right)
}\right) $ satisfy $w_{k}\in \left( 0,1\right) $ and $\sum_{k\geq 1}w_{k}=1.$
With $\varphi _{\xi }\left( q\right) :={\Bbb E}\xi ^{q}$, its moment
function is equivalently given by 
\begin{equation}
\varphi _{\xi }\left( q\right) =\frac{1}{\theta }\sum_{k\geq 1}{\Bbb E}%
\left( \xi _{\left( k\right) }\right) \varphi _{\xi _{\left( k\right)
}}\left( q\right)  \label{eq18}
\end{equation}
in terms of $\varphi _{\xi _{\left( k\right) }}\left( q\right) :={\Bbb E}\xi
_{\left( k\right) }^{q},$ the moment functions of $\xi _{\left( k\right) }.$%
\newline

\begin{center}
{\bf Size-biased Picking from} $\left( \xi _{\left( k\right) },k\geq
1\right) $
\end{center}

Let $\eta $ be a $\left( 0,1\right) -$valued random variable taking the
value $\xi _{\left( k\right) }$ with probability $\frac{1}{\theta }\xi
_{\left( k\right) }$ given $\left( \xi _{\left( k\right) },k\geq 1\right) $.
This random variable corresponds to a size-biased picking from $\left( \xi
_{\left( k\right) },k\geq 1\right) $. Its moment function is 
\begin{equation}
\varphi _{\eta }\left( q\right) ={\Bbb E}\eta ^{q}:={\Bbb E}\frac{1}{\theta }%
\sum_{k\geq 1}\xi _{\left( k\right) }\xi _{\left( k\right) }^{q}=\frac{1}{%
\theta }\phi \left( q+1\right)  \label{eq19}
\end{equation}
for $q>q_{*}:=\beta _{*}-1\in \left[ -1,0\right) .$

The waiting time paradox reads 
\begin{equation}
\eta \succeq _{st}\xi ,  \label{eq20}
\end{equation}
a stochastic domination property translating the fact that in the
size-biased picking procedure, large fragments are favored.\newline

\subsection{Spacings and Strong Partition of Unity: Normalizing}

\indent

The partition $\left( \xi _{\left( k\right) };\text{ }k\geq 1\right) $ of $%
\chi $ induces another natural partition of unity defined as follows. Define
the incremental random variables $\widetilde{\xi }_{k}:=\xi _{\left(
k-1\right) }-\xi _{\left( k\right) }$ (with $\xi _{\left( 0\right) }:=1$), $%
k\geq 1$. Then, $\left( \widetilde{\xi }_{k},k\geq 1\right) $ defines a new
sequence of $\left( 0,1\right) -$valued random variables with clearly $%
\sum_{k\geq 1}\widetilde{\xi }_{k}=1$ (almost surely). The $\left( 
\widetilde{\xi }_{k};\text{ }k\geq 1\right) $ constitute a strong (almost
sure) random partition of unity built on $\chi $. Spacings between
consecutive ordered energies sum up to $1$, which is the top energy a single
event can develop according to our assumptions. This model was first
considered by \cite{Ignatov} and reconsidered by \cite{Arr} in the context
of combinatorial structures.\newline

That this construction is possible is indeed a consequence of $\Pi $ being
concentrated on $\left( 0,1\right) $ leading to $\left( 0,1\right) -$valued $%
\xi _{\left( k\right) }$s. Note that there are no reasons, in general, for
the $\widetilde{\xi }_{k}$s to be ordered either in the strict or weaker
stochastic sense. The ordered version of $\left( \widetilde{\xi }_{k};\text{ 
}k\geq 1\right) $, say $\left( \widetilde{\xi }_{\left( k\right) };\text{ }%
k\geq 1\right) ,$ is thus also of some interest.

This construction should not be confused with another partition of unity
which can be defined from the system of ordered normalized random weights $%
\varsigma _{\left( k\right) }:=\xi _{\left( k\right) }/\chi $, $k\geq 1$
satisfying $\sum_{k\geq 1}\varsigma _{\left( k\right) }\stackrel{d}{=}1$ and 
$\varsigma _{\left( 1\right) }>..>\varsigma _{\left( k\right) }>..$. For
this kind of partition of unity, the condition that L\'{e}vy measure $\Pi $
of $\chi $ be concentrated on $\left( 0,1\right) $ is inessential. When $\Pi 
$ is concentrated on $\left( 0,\infty \right) $, one speaks of
Poisson-Kingman partitions (see \cite{Pit4}). For instance, when $\Pi \left(
dx\right) =\theta e^{-x}/x,$ $x>0$, is the L\'{e}vy measure for jumps of a
Moran (gamma) subordinator $\left( \chi _{t};t\geq 0\right) $, $\left(
\varsigma _{\left( k\right) };\text{ }k\geq 1\right) $ has Poisson-Dirichlet
PD$\left( \theta \right) $ distribution, independent of $\chi =\chi _{1}$.
For such problems, the joint law of $\left( \chi ;\varsigma _{\left(
k\right) }\text{, }k=1,..,l\right) $ for each $l\geq 1$ deserves some
attention. They are given by Perman formulae (see \cite{Perman}, for
additional details).\newline

\begin{center}
{\bf Strong Partition Function of Unity from} $\chi ${\bf : Structural
Measure}
\end{center}

Let 
\[
\widetilde{\phi }\left( \beta \right) :={\Bbb E}\sum_{k\geq 1}\widetilde{\xi 
}_{k}^{\beta }\text{, with }\beta >\beta _{*}\in \left[ 0,1\right) . 
\]
With $S_{0}:=0$, averaging over $S_{k}\stackrel{d}{\sim }$ gamma$\left(
k\right) $ , $k\geq 1$, $\widetilde{\phi }\left( \beta \right) $ can be
obtained in general from 
\begin{equation}
\widetilde{\phi }\left( \beta \right) =\sum_{k\geq 1}\int_{0}^{\infty }e^{-t}%
{\Bbb E}\left[ \left( \overline{\Pi }^{-1}\left( S_{k-1}\right) -\overline{%
\Pi }^{-1}\left( S_{k-1}+t\right) \right) ^{\beta }\right] dt,  \label{eq21}
\end{equation}
recalling $S_{k}=S_{k-1}+T_{k}$ where $S_{k-1}$ is independent of $T_{k}%
\stackrel{d}{\sim }$ exp$\left( 1\right) $. The measure $\sigma \left(
dx\right) $ such that $\widetilde{\phi }\left( \beta \right)
=\int_{0}^{1}x^{\beta }\sigma \left( dx\right) $ is called the structural
measure of the partition $\left( \widetilde{\xi }_{k},k\geq 1\right) $. With 
$\overline{\sigma }\left( x\right) :=\int_{x}^{1}\sigma \left( dz\right) $,
recalling ${\Bbb P}\left( S_{k}\in ds\right) =\frac{1}{\left( k-1\right) !}%
s^{k-1}e^{-s}ds$, it can generally be obtained, after a change of variable,
from 
\[
\overline{\sigma }\left( x\right) =\sum_{k\geq 1}\int_{0}^{\infty }e^{-t}%
{\Bbb P}\left[ \overline{\Pi }^{-1}\left( S_{k-1}\right) -\overline{\Pi }%
^{-1}\left( S_{k-1}+t\right) >x\right] dt 
\]
\[
=e^{-\overline{\Pi }\left( 1-x\right) }+\sum_{k\geq 1}{\Bbb E}\left\{
e^{-\left[ \overline{\Pi }\left( \overline{\Pi }^{-1}\left( S_{k}\right)
-x\right) -S_{k}\right] };\text{ }S_{k}<\overline{\Pi }\left( x\right)
\right\} 
\]
\[
=e^{-\overline{\Pi }\left( 1-x\right) }-\sum_{k\geq 1}\frac{1}{k!}%
\int_{x}^{1}e^{-\overline{\Pi }\left( z-x\right) }d\overline{\Pi }^{k}\left(
z\right) 
\]
\begin{equation}
=e^{-\overline{\Pi }\left( 1-x\right) }+\int_{x}^{1}e^{-\left[ \overline{\Pi 
}\left( z-x\right) -\overline{\Pi }\left( z\right) \right] }\pi \left(
z\right) dz.  \label{eq21a}
\end{equation}
The random variable $\sum_{k\geq 1}\widetilde{\xi }_{k}^{\beta }$ is called
the partition function of unity constructed from $\chi $ and $\sigma $
defined by $\overline{\sigma }\left( x\right) =\sum_{k\geq 1}{\Bbb P}\left( 
\widetilde{\xi }_{k}>x\right) $ its structural measure.\newline

\begin{center}
{\bf Cutoff Considerations for Spacings Partition}
\end{center}

1/ Overshoot. First, we note that, if $\overline{\xi }_{k}:=1-\sum_{l=1}^{k}%
\widetilde{\xi }_{l}$ is the amount of space left vacant by the $k$ first
atoms of $\left( \widetilde{\xi }_{k},k\geq 1\right) $, we get $\overline{%
\xi }_{k}=\xi _{\left( k\right) }.$ This remark allows us to derive the
following result.

Let $x\in \left( 0,1\right) $ be some threshold value. Define 
\[
K\left( x\right) :=\inf \left( k\geq 1:\sum_{l=1}^{k}\widetilde{\xi }%
_{l}>x\right) 
\]
to be the first time the cumulated fragments size of $\left( \widetilde{\xi }%
_{k},k\geq 1\right) $ exceeds $x.$ Then, 
\[
K\left( x\right) \stackrel{d}{=}1+P_{\Lambda \left( x\right) } 
\]
where $P_{\Lambda \left( x\right) }$ is a Poisson distributed random
variable with parameter $\Lambda \left( x\right) :=\overline{\Pi }\left(
1-x\right) .$

Indeed, ${\Bbb P}\left( K\left( x\right) >k\right) ={\Bbb P}\left(
\sum_{l=1}^{k}\widetilde{\xi }_{l}\leq x\right) ={\Bbb P}\left( \xi _{\left(
k\right) }>1-x\right) .$

The random quantity $\sum_{l=1}^{K\left( x\right) }\widetilde{\xi }_{l}-x$
is the overshoot at $x.$ \newline

2/ Let ${\cal N}_{+}\left( \epsilon \right) :=\sum_{k\geq 1}{\bf I}\left( 
\widetilde{\xi }_{k}>\epsilon \right) $ be the random number of spacings $%
\widetilde{\xi }_{k}$ exceeding $\epsilon \in \left( 0,1\right) $. Then,
from the above expression of $\overline{\sigma }\left( x\right) $ in Eq. (%
\ref{eq21a}), evaluated in a neighborhood of $x=0$, 
\[
{\Bbb E}{\cal N}_{+}\left( \epsilon \right) =\overline{\sigma }\left(
\epsilon \right) \sim _{\epsilon \downarrow 0}\overline{\Pi }\left( \epsilon
\right) , 
\]
and Chen-Stein methods for Poisson approximations of ${\cal N}_{+}\left(
\epsilon \right) $ could be developed, in the spirit of \cite{Hirth}.

However, ${\Bbb P}\left( \inf \left( l\geq 1:\widetilde{\xi }_{l}\leq
\epsilon \right) >k\right) ={\Bbb P}\left( \wedge _{l=1}^{k}\widetilde{\xi }%
_{k}>\epsilon \right) $ where $\wedge _{l=1}^{k}\widetilde{\xi }_{l}$ is the
smallest term amongst $\left( \widetilde{\xi }_{1},..,\widetilde{\xi }%
_{k}\right) $ and it is no longer true that ${\cal N}_{+}\left( \epsilon
\right) =\inf \left( k\geq 1:\widetilde{\xi }_{k}\leq \epsilon \right) -1$
because the $\widetilde{\xi }_{k}$ are not ordered.

The contribution to total mass of those $\widetilde{\xi }_{k}$ above or
below $\epsilon $ are respectively $1_{+}\left( \epsilon \right)
:=\sum_{k\geq 1}\widetilde{\xi }_{k}{\bf I}\left( \widetilde{\xi }%
_{k}>\epsilon \right) $ and $1_{-}\left( \epsilon \right) :=\sum_{k\geq 1}%
\widetilde{\xi }_{k}{\bf I}\left( \widetilde{\xi }_{k}\leq \epsilon \right) $%
. The full laws of these quantities are difficult to obtain in general.
Indeed, 
\[
{\Bbb E}\exp \left\{ -\lambda {\cal N}_{+}\left( \epsilon \right) \right\} =%
{\Bbb E}\prod_{k\geq 1}\left( 1-\left( 1-e^{-\lambda }\right) {\bf I}\left( 
\widetilde{\xi }_{k}>\epsilon \right) \right) 
\]
\[
{\Bbb E}\exp \left\{ -\lambda 1_{\pm }\left( \epsilon \right) \right\} =%
{\Bbb E}\prod_{k\geq 1}\left( 1-\left( 1-e^{-\lambda \widetilde{\xi }%
_{k}}\right) {\bf I}\left( \widetilde{\xi }_{k}\QTATOP{>}{\leq }\epsilon
\right) \right) 
\]
and the joint laws of the $\widetilde{\xi }_{k}$ are required. However, as $%
\epsilon \downarrow 0$, it still holds that 
\[
1_{+}\left( \epsilon \right) \stackrel{d}{\rightarrow }1\text{ and }{\Bbb E}%
1_{-}\left( \epsilon \right) \sim \epsilon ^{2}\sum_{k\geq 1}f_{\widetilde{%
\xi }_{k}}\left( \epsilon \right) =\epsilon ^{2}\sigma \left( \epsilon
\right) \sim \epsilon ^{2}\pi \left( \epsilon \right) . 
\]
\newline

\begin{center}
{\bf Size-biased Picking from} $\left( \widetilde{\xi }_{k},k\geq 1\right) $
\end{center}

Let $\widetilde{\eta }$ be a $\left( 0,1\right) -$valued random variable
taking the value $\widetilde{\xi }_{k}$ with probability $\widetilde{\xi }%
_{k}$ given $\left( \widetilde{\xi }_{k},k\geq 1\right) $. This random
variable corresponds to a size-biased picking from $\left( \widetilde{\xi }%
_{k},k\geq 1\right) $. Its moment function is 
\begin{equation}
\varphi _{\widetilde{\eta }}\left( q\right) ={\Bbb E}\widetilde{\eta }^{q}:=%
{\Bbb E}\sum_{k\geq 1}\widetilde{\xi }_{k}^{q+1}=\widetilde{\phi }\left(
q+1\right)  \label{eq23}
\end{equation}
for $q>q_{*}:=\beta _{*}-1\in \left[ -1,0\right) .$\newline

Just like for the pair $\eta $ and $\xi $, the waiting time paradox reads $%
\widetilde{\eta }\succeq _{st}\widetilde{\xi }$.\newline

\section{Examples: Dickman partition and related ones}

\indent

We start with a fundamental example in many respects, for which most
computations can be painlessly achieved. We call it Dickman partition for
reasons to appear later. The peculiarities of this model clearly appeared in
the Examples developed to illustrate the general partition model under study
in the previous Sections.

\subsection{Dickman Partition}

$\bullet $ Assume $\Pi \left( dx\right) =\frac{\theta }{x}{\bf I}_{x\in
\left( 0,1\right) }dx.$ Then $\overline{\Pi }\left( x\right) =-\theta \log x$
and $\overline{\Pi }^{-1}\left( s\right) =e^{-s/\theta }$. The LST of $\chi $
in this case is 
\begin{eqnarray*}
{\Bbb E}e^{-\lambda \chi } &=&\exp \left\{ -\theta \int_{0}^{1}\frac{%
1-e^{-\lambda x}}{x}dx\right\} \\
&=&\exp \left\{ -\theta \int_{0}^{\lambda }\frac{1-e^{-\lambda ^{\prime }}}{%
\lambda ^{\prime }}d\lambda ^{\prime }\right\} .
\end{eqnarray*}
Let us now proceed with the detailed study of the multiplicative structure
of this partitioning model.\newline

\begin{center}
{\bf Partition Function}
\end{center}

In this example, we have $\xi _{\left( k\right) }=e^{-S_{k}/\theta
}=\prod_{l=1}^{k}B_{l}$ where $B_{k}$ are iid with beta$\left( \theta
,1\right) $ law: ${\Bbb P}\left( B_{1}\leq x\right) =x^{\theta }.$ The $\xi
_{\left( k\right) }$ are thus log-gamma$\left( k,\theta \right) $
distributed. We obtain 
\begin{equation}
{\Bbb E}\xi _{\left( k\right) }^{\beta }=\left( \frac{\theta }{\theta +\beta 
}\right) ^{k}\text{ and }{\Bbb E}\sum_{k\geq 1}\xi _{\left( k\right)
}^{\beta }=\frac{\theta }{\beta }=\int_{0}^{1}x^{\beta }\frac{\theta }{x}dx,
\label{eq24}
\end{equation}
for $\beta >\beta _{*}=0.$ The structural measure is $\sigma \left(
dx\right) =\frac{\theta }{x}dx.$

Note also that $-\frac{\theta }{k}\log \left( \xi _{\left( k\right) }\right)
\rightarrow 1$ almost surely as $k\uparrow \infty $ so that $\xi _{\left(
k\right) }$ goes to $0$ exponentially fast with $k$.

Note that, exploiting the product decomposition of the $\xi _{\left(
k\right) }$, if $\varsigma _{\left( 1\right) }:=\xi _{\left( 1\right) }/\chi 
$ is the largest normalized fragment of the $\xi _{\left( k\right) }$'s, we
get $\varsigma _{\left( 1\right) }\stackrel{d}{=}1/\left( 1+\chi \right) $.
Consequently, 
\begin{equation}
{\Bbb E}e^{-\lambda /\varsigma _{\left( 1\right) }}=\exp \left\{ -\lambda
-\theta \int_{0}^{1}\frac{1-e^{-\lambda x}}{x}dx\right\}  \label{eq25}
\end{equation}
is the infinitely divisible LST of $1/\varsigma _{\left( 1\right) }>1.$ 
\newline

\begin{center}
{\bf Correlation Structure}
\end{center}

The correlation structure of the Dickman partition has already been computed
in a former example. The result is the expression of $C_{l}\left(
q_{1},q_{2}\right) $ in (\ref{e2}).\newline

\begin{center}
{\bf Typical and Size-biased Fragment Size}
\end{center}

The size-biased picking random variable $\eta $ from $\left( \xi _{\left(
k\right) };k\geq 1\right) $ has uniform law since 
\begin{equation}
{\Bbb E}\eta ^{q}:={\Bbb E}\frac{1}{\theta }\sum_{k\geq 1}\xi _{\left(
k\right) }^{q+1}=1/\left( q+1\right) .  \label{eq26}
\end{equation}
The typical fragment size $\xi $ has law given by 
\begin{equation}
\varphi _{\xi }\left( q\right) ={\Bbb E}\xi ^{q}:=\frac{1}{\theta }%
\sum_{k\geq 1}{\Bbb E}\xi _{\left( k\right) }{\Bbb E}\xi _{\left( k\right)
}^{q}=\frac{\theta }{\left( 1+\theta \right) q+\theta }  \label{eq27}
\end{equation}
corresponding to a beta$\left( \frac{\theta }{1+\theta },1\right) $
distribution. It is true that $\eta \succeq _{st}\xi $ since $F_{\eta
}\left( x\right) =x\leq F_{\xi }\left( x\right) =x^{\frac{\theta }{1+\theta }%
}$ for all $x\in \left[ 0,1\right] .$\newline

\begin{center}
{\bf Additional Properties: the Law of }$\chi $
\end{center}

With $\gamma $ the Euler constant, the random variable $\chi $ has a density
given by 
\begin{equation}
f_{\chi }\left( x\right) =e^{-\gamma \theta }x^{\theta -1}\overline{F}%
_{\theta }\left( x\right) /\Gamma \left( \theta \right) \text{, }x>0
\label{eq28}
\end{equation}
where $\overline{F}_{\theta }\left( x\right) :={\Bbb P}\left( D>x\right) $
is the tail probability distribution function of Dickman random variable $D$%
\[
\overline{F}_{\theta }\left( x\right) ={\bf I}_{x\in \left[ 0,1\right) }+%
{\bf I}_{x\geq 1}\left( 1+\sum_{j=1}^{\left[ x\right] }\frac{\left( -\theta
\right) ^{j}}{j!}\int_{\frac{1}{x}}^{1}..\int_{\frac{1}{x}}^{1}\frac{\left(
1-\sum_{l=1}^{j}z_{l}\right) _{+}^{\theta -1}}{\prod_{l=1}^{j}z_{l}}%
\prod_{l=1}^{j}dz_{l}\right) 
\]
with super-exponential von Mises tails $-\log \overline{F}_{\theta }\left(
x\right) \sim _{x\uparrow \infty }x\log x$ (see \cite{Holstb}). When $x\geq
1 $, the function $\overline{F}_{\theta }\left( x\right) $ is the solution
to 
\[
x^{\theta }\overline{F}_{\theta }\left( x\right) =\theta
\int_{x-1}^{x}z^{\theta -1}\overline{F}_{\theta }\left( z\right) dz. 
\]
The random variable $D>1$ turns out to be the reciprocal of the largest
normalized jump of the Moran (gamma) subordinator ($D=1/\varsigma _{\left(
1\right) }$).

This relationship between $f_{\chi }\left( x\right) $ and $\overline{F}%
_{\theta }\left( x\right) $ may be seen to be a direct consequence of the
identity 
\[
\exp \left( -\theta \left( \int_{1}^{\infty }\frac{e^{-\lambda x}}{x}dx+\log
\lambda +\gamma \right) \right) =\exp \left( -\theta \int_{0}^{1}\frac{%
1-e^{-\lambda x}}{x}dx\right) 
\]
involving the exponential integral function $\func{Ei}\left( \lambda \right)
=\int_{1}^{\infty }\frac{e^{-\lambda x}}{x}dx$ (see \cite{Holstb}). For
these connections with Dickman's function, we shall say from (\ref{eq28})
that $\chi $ follows Dickman distribution. \newline

\begin{center}
{\bf Moment Function of} $\chi $
\end{center}

The moment function of $\chi $, say ${\Bbb E}\chi ^{q}=\int_{0}^{\infty
}x^{q}f_{\chi }\left( x\right) dx$, is defined for $q>-\theta $, with 
\[
{\Bbb E}\chi ^{q}=\frac{e^{-\gamma \theta }}{\Gamma \left( \theta \right)
\left( q+\theta \right) }{\Bbb E}D^{q+\theta } 
\]
It can be computed as follows: first, $\chi \stackrel{d}{=}\xi _{\left(
1\right) }\left( 1+\chi ^{^{\prime }}\right) $ where $\chi ^{^{\prime }}%
\stackrel{d}{=}\chi $ is independent of $\xi _{\left( 1\right) }\stackrel{d}{%
\sim }$ beta$\left( \theta ,1\right) $ with ${\Bbb E}\xi _{\left( 1\right)
}^{q}=\theta /\left( \theta +q\right) $. Thus, $\chi $ is a Vervaat
perpetuity of a special type.

Let $\mu _{n}\left( \theta \right) :={\Bbb E}\xi _{\left( 1\right) }^{n}$;
from this, using the binomial identity, the integral moments $m_{n}\left(
\theta \right) $ of $\chi $ are first obtained recursively by $m_{0}\left(
\theta \right) =1$ and 
\begin{equation}
m_{n}\left( \theta \right) =\frac{\mu _{n}\left( \theta \right) }{1-\mu
_{n}\left( \theta \right) }\sum_{p=0}^{n-1}\binom{n}{p}m_{p}\left( \theta
\right) \text{, }n\geq 1,  \label{eq29}
\end{equation}
with $\frac{\mu _{n}\left( \theta \right) }{1-\mu _{n}\left( \theta \right) }%
=\frac{\theta }{n}$. Hence 
\begin{eqnarray*}
m_{1}\left( \theta \right) &=&\theta , \\
m_{2}\left( \theta \right) &=&\frac{\theta }{2}\left( 1+2m_{1}\left( \theta
\right) \right) =\frac{\theta }{2}+\theta ^{2}, \\
m_{3}\left( \theta \right) &=&\frac{\theta }{3}\left( 1+3m_{1}\left( \theta
\right) +3m_{2}\left( \theta \right) \right) =\frac{\theta }{3}+\frac{%
3\theta ^{2}}{2}+\theta ^{3}
\end{eqnarray*}
are the three first nested moments. Searching for solutions under the
polynomial form 
\[
m_{n}\left( \theta \right) =\sum_{k=1}^{n}b_{k,n}\theta ^{k}\text{, }n\geq 1 
\]
we can identify the coefficients $b_{k,n}$, $k=2,..,n$ as the ones solving ($%
b_{1,n}=1/n$) the Bell numbers-like triangular recurrence 
\[
b_{k,n}=\frac{1}{n}\sum_{p=k-1}^{n-1}\binom{n}{p}b_{k-1,p}\text{, }k=2,..,n. 
\]
Next, with $\left( q\right) _{n}:=q\left( q-1\right) ..\left( q-n+1\right) $%
, the full expression of ${\Bbb E}\chi ^{q}$ is 
\begin{equation}
{\Bbb E}\chi ^{q}=\frac{\theta }{\theta +q}\left( 1+\sum_{n\geq 1}\frac{%
\left( q\right) _{n}}{n!}m_{n}\left( \theta \right) \right) ,\text{ }%
q>-\theta ,  \label{eq30}
\end{equation}
translating the identity ${\Bbb E}\chi ^{q}=\frac{\theta }{\theta +q}{\Bbb E}%
\left( 1+\chi \right) ^{q}.$ Incidentally, 
\[
{\Bbb E}D^{q}=\Gamma \left( \theta +1\right) e^{\gamma \theta }\left(
1+\sum_{n\geq 1}\frac{\left( q-\theta \right) _{n}}{n!}m_{n}\left( \theta
\right) \right) ,q>0 
\]
is the moment function of $D=1/\varsigma _{\left( 1\right) }.$

Let $\chi _{1/\beta }:=\sum_{k\geq 1}\xi _{\left( k\right) }^{\beta }$, with 
$\beta >0$. From Campbell formula, 
\begin{equation}
{\Bbb E}e^{-\lambda \chi _{1/\beta }}=\exp \left\{ -\frac{1}{\beta }%
\int_{0}^{1}\left( 1-e^{-\lambda x}\right) \Pi \left( dx\right) \right\}
=\left( {\Bbb E}e^{-\lambda \chi }\right) ^{1/\beta }.  \label{eq31}
\end{equation}
The moment function ${\Bbb E}\chi _{1/\beta }^{q}$ can thus readily be
obtained from the one of ${\Bbb E}\chi ^{q}$ while operating the
substitution $\theta \rightarrow \theta /\beta $ in the obtained formula.
Hence 
\begin{equation}
{\Bbb E}\chi _{1/\beta }^{q}=\frac{\theta }{\theta +\beta q}\left(
1+\sum_{n\geq 1}\frac{\left( q\right) _{n}}{n!}m_{n}\left( \theta /\beta
\right) \right) ,\text{ }q>-\theta /\beta  \label{eq32}
\end{equation}
is the moment function of the partition function $\sum_{k\geq 1}\xi _{\left(
k\right) }^{\beta }.$ Note that ${\Bbb E}\chi _{1/\beta }=\frac{\theta }{%
\theta +\beta }\left( 1+\theta /\beta \right) =\theta /\beta $, as required.
This constitutes a complementary information to the one encoded in the LST
of $\sum_{k\geq 1}\xi _{\left( k\right) }^{\beta }$. This suggests the
following additional construction.\newline

\begin{center}
{\bf R\'{e}nyi's} {\bf Weighted Averages}
\end{center}

Let $\varsigma _{\left( k\right) }:=\xi _{\left( k\right) }/\chi $, $k\geq 1$
be a system of normalized random weights. With $\beta >-1$, define the
random R\'{e}nyi $\beta -$average $\left[ \xi \right] _{\beta }$ (with
random weights $\varsigma _{\left( k\right) }$) of the $\left( \xi _{\left(
k\right) },k\geq 1\right) $ to be ($\beta >-1$) 
\[
\left[ \xi \right] _{\beta }:=\left( \sum_{k\geq 1}\varsigma _{\left(
k\right) }\xi _{\left( k\right) }^{\beta }\right) ^{1/\beta }=\left( \frac{1%
}{\chi }\sum_{k\geq 1}\xi _{\left( k\right) }^{\beta +1}\right) ^{1/\beta
}=\left( \frac{\chi _{1/\left( \beta +1\right) }}{\chi }\right) ^{1/\beta }. 
\]
This random variable is $\left( 0,1\right) -$valued when $\beta >-1$ and
when $\beta \leq -1$, it degenerates to $0$. The $2-$average $\left\langle
\xi \right\rangle _{2}$ is often considered, but $\left[ \xi \right]
_{0}:=\lim_{\beta \uparrow 0}\left[ \xi \right] _{\beta }=\prod_{k\geq 1}\xi
_{\left( k\right) }^{\varsigma _{k}}$ is also sometimes of interest. The
function $\beta \rightarrow \left[ \xi \right] _{\beta }$ is non-decreasing
with $\beta $ and $\left[ \xi \right] _{\beta _{1}}>\left[ \xi \right]
_{\beta _{2}}$ if $-1<\beta _{2}<\beta _{1}$.

Note that ${\Bbb E}\left[ \xi \right] _{\beta }^{q}$ is also ${\Bbb E}\left[
e^{-qH_{\beta }}\right] $ where $H_{\beta }=-\log \left[ \xi \right] _{\beta
}$ is the random R\'{e}nyi $\beta -$entropy of the sequence $\left( \xi
_{\left( 1\right) },..,\xi _{\left( k\right) },..\right) $, with, in
particular, $H_{0}=-\log \left[ \xi \right] _{0}=-\sum_{k\geq 1}\varsigma
_{k}\log \xi _{\left( k\right) }$, related to Shannon entropy.

The computation of its moment function turns out to be difficult as it
stands. Indeed, recalling that $\left( \chi _{t},t\geq 0\right) $ is a
process with stationary independent increments, the joint moment function of 
$\chi _{1/\left( \beta +1\right) }$ and $\chi =\chi _{1}$ would first be
required.\newline

We shall rather consider the simpler R\'{e}nyi $\beta -$average (with
deterministic weights $w_{k}$) of the $\left( \xi _{\left( k\right) },k\geq
1\right) $ to be 
\begin{equation}
\left\langle \xi \right\rangle _{\beta }:=\left( \sum_{k\geq 1}w_{k}\xi
_{\left( k\right) }^{\beta }\right) ^{1/\beta }\text{ where }w_{k}=\frac{%
{\Bbb E}\xi _{\left( k\right) }}{\theta }\text{ and }\beta >0.  \label{eq33}
\end{equation}
It may be checked that the range of the random variable $\left\langle \xi
\right\rangle _{\beta }$ is $\left[ 0,1\right] $ when $\beta \in \left(
0,\infty \right) $ which we shall limit ourselves to. Define the weighted
sum 
\[
W:=\sum_{k\geq 1}w_{k}\xi _{\left( k\right) }^{\beta } 
\]
Recalling $w_{k}=\frac{1}{\theta }\left( \frac{\theta }{\theta +1}\right)
^{k}$ and $\xi _{\left( k\right) }=\prod_{l=1}^{k}B_{k}$ where $B_{k}$ are
iid with $\xi _{\left( 1\right) }=B_{1}\stackrel{d}{\sim }$ beta$\left(
\theta ,1\right) $ law, we have 
\[
W\stackrel{d}{=}\frac{\xi _{\left( 1\right) }^{\beta }}{\theta +1}\left(
1+\theta W^{\prime }\right) 
\]
where $W^{\prime }\stackrel{d}{=}W$ is independent of $\xi _{\left( 1\right)
}^{\beta }\stackrel{d}{\sim }$ beta$\left( \frac{\theta }{\beta },1\right) $%
. Let $\mu _{n}\left( \theta ,\beta \right) :=\frac{1}{\left( \theta
+1\right) ^{n}}{\Bbb E}\xi _{\left( 1\right) }^{n\beta }=\frac{1}{\left(
\theta +1\right) ^{n}}\frac{\theta }{\theta +n\beta }$; from this, using the
binomial identity, the integral moments $m_{n}\left( \theta ,\beta \right) $
of $W$ are first obtained recursively by $m_{0}\left( \theta ,\beta \right)
=1$ and 
\[
m_{n}\left( \theta ,\beta \right) =\frac{\mu _{n}\left( \theta ,\beta
\right) }{1-\mu _{n}\left( \theta ,\beta \right) }\sum_{p=0}^{n-1}\binom{n}{p%
}\theta ^{p}m_{p}\left( \theta ,\beta \right) \text{, }n\geq 1, 
\]
with $\frac{\mu _{n}\left( \theta \right) }{1-\mu _{n}\left( \theta \right) }%
=\frac{\theta }{\left( \theta +1\right) ^{n}\left( \theta +n\beta \right)
-\theta }$. From this 
\begin{equation}
{\Bbb E}W^{q}=\theta \frac{1+\sum_{n\geq 1}\frac{\left( q\right) _{n}}{n!}%
\theta ^{n}m_{n}\left( \theta ,\beta \right) }{\left( 1+\theta \right)
^{q}\left( \theta +\beta q\right) }  \label{eq34}
\end{equation}
and 
\begin{equation}
{\Bbb E}\left\langle \xi \right\rangle _{\beta }^{q}={\Bbb E}W^{q/\beta
}=\theta \frac{1+\sum_{n\geq 1}\frac{\left( q/\beta \right) _{n}}{n!}\theta
^{n}m_{n}\left( \theta ,\beta \right) }{\left( 1+\theta \right) ^{q/\beta
}\left( \theta +q\right) }.  \label{eq35}
\end{equation}
\newline

\subsection{Spacings: an Alternative Construction of Poisson-Dirichlet
Partition}

\indent

Defining spacings to be $\widetilde{\xi }_{k}:=\xi _{\left( k-1\right) }-\xi
_{\left( k\right) }$ (with $\xi _{\left( 0\right) }:=1$), we clearly have 
\begin{equation}
\widetilde{\xi }_{k}\stackrel{d}{=}\prod_{l=1}^{k-1}\overline{v}_{l}v_{k}%
\text{, }k\geq 1  \label{eq36}
\end{equation}
where $v_{k}$ are iid with beta$\left( 1,\theta \right) $ law: ${\Bbb P}%
\left( v_{1}\leq x\right) =\left( 1-x\right) ^{\theta }$. Thus $\left( 
\widetilde{\xi }_{k},k\geq 1\right) $ has GEM$\left( \theta \right) $
distribution with $\sum_{k\geq 1}\widetilde{\xi }_{k}=1$.

In this particular example, one can also check that $\widetilde{\xi }%
_{1}\succeq _{st}..\succeq _{st}\widetilde{\xi }_{k}\succeq _{st}..$ and the 
$\left( \widetilde{\xi }_{k},k\geq 1\right) $ are arranged in stochastic
descending order. Finally, as is well-known, the ordered version $\left( 
\widetilde{\xi }_{\left( k\right) },k\geq 1\right) $ of $\left( \widetilde{%
\xi }_{k},k\geq 1\right) $ has Poisson-Dirichlet PD$\left( \theta \right) -$%
distribution and $\left( \widetilde{\xi }_{k},k\geq 1\right) $, as a
size-biased permutation from PD$\left( \theta \right) $, is invariant under
size-biased permutation.\newline

\begin{center}
{\bf Correlation Structure}
\end{center}

With $w_{k}:={\Bbb E}\widetilde{\xi }_{k}$, the second order quantity to
consider here is 
\begin{equation}
\widetilde{C}_{l}\left( q_{1},q_{2}\right) :={\Bbb E}\left( \sum_{k\geq
1}w_{k}\widetilde{\xi }_{k}^{q_{1}}\widetilde{\xi }_{k+l}^{q_{2}}\right) .
\label{e8}
\end{equation}
From the multiplicative RAM structure of the GEM$\left( \theta \right) $
partition, 
\[
\widetilde{C}_{l}\left( q_{1},q_{2}\right) =\frac{1}{1+\theta }\sum_{k\geq
1}\left( \frac{\theta }{1+\theta }\right) ^{k-1}{\Bbb E}\left(
\prod_{l=1}^{k-1}\overline{v}_{l}^{q_{1}+q_{2}}v_{k}^{q_{1}}\overline{v}%
_{k}^{q_{2}}\prod_{l=k+1}^{k+l-1}\overline{v}_{l}^{q_{2}}v_{k+l}^{q_{2}}%
\right) 
\]
\[
=\sum_{k\geq 1}\left( \frac{\theta }{1+\theta }\right) ^{k}\left( \frac{%
\theta }{\theta +q_{1}+q_{2}}\right) ^{k-1}\frac{\Gamma \left(
1+q_{1}\right) \Gamma \left( \theta +q_{2}\right) }{\Gamma \left( 1+\theta
+q_{1}+q_{2}\right) }\times 
\]
\[
\left( \frac{\theta }{\theta +q_{2}}\right) ^{l-1}\left( \frac{\Gamma \left(
1+q_{2}\right) \Gamma \left( 1+\theta \right) }{\Gamma \left( 1+\theta
+q_{2}\right) }\right) 
\]
\begin{equation}
=K\left( q_{1},q_{2}\right) \left( \frac{\theta }{\theta +q_{2}}\right) ^{l}
\label{e9}
\end{equation}
where 
\begin{equation}
K\left( q_{1},q_{2}\right) =\frac{\Gamma \left( 1+q_{1}\right) \Gamma \left(
1+q_{2}\right) \Gamma \left( 1+\theta \right) }{\Gamma \left( \theta
+q_{1}+q_{2}\right) \left[ \theta +\left( \theta +1\right) \left(
q_{1}+q_{2}\right) \right] }  \label{e10}
\end{equation}
is defined for $\left\{ q_{1}+q_{2}>-\theta /\left( \theta +1\right) ;\left(
q_{1},q_{2}\right) >-1\right\} .$ The definition domain of $\widetilde{C}%
_{l}\left( q_{1},q_{2}\right) $ therefore is $\left\{ q_{1}+q_{2}>-\theta
/\left( \theta +1\right) ;q_{1}>-1;q_{2}>-\min \left( 1,\theta \right)
\right\} .$ This should be compared with the expression of $C_{l}\left(
q_{1},q_{2}\right) $ in (\ref{e2}), dealing with the unnormalized case.%
\newline

\begin{center}
{\bf Cutoff Considerations for GEM}$\left( \theta \right) ${\bf \ Partition}
\end{center}

Let ${\cal N}_{+}\left( \epsilon \right) :=\sum_{k\geq 1}{\bf I}\left( 
\widetilde{\xi }_{k}>\epsilon \right) $ be the random number of spacings $%
\widetilde{\xi }_{k}$ exceeding $\epsilon \in \left( 0,1\right) $. Then,
using the RAM structure of $\left( \widetilde{\xi }_{k};k\geq 1\right) $, we
have 
\[
{\cal N}_{+}\left( \epsilon \right) \stackrel{d}{=}{\bf I}\left(
v_{1}>\epsilon \right) +{\cal N}_{+}^{^{\prime }}\left( \epsilon /\overline{v%
}_{1}\right) {\bf I}_{\overline{v}_{1}>\epsilon } 
\]
where ${\cal N}_{+}^{^{\prime }}\left( .\right) $ is a statistical copy of $%
{\cal N}_{+}\left( .\right) .$ Here $v_{1}\stackrel{d}{\sim }$ beta$\left(
1,\theta \right) $ and $\overline{v}_{1}:=1-v_{1}$ is independent of ${\cal N%
}_{+}^{^{\prime }}\left( .\right) $. In particular, if $n_{+}^{\left(
p\right) }\left( \epsilon \right) ={\Bbb E}{\cal N}_{+}\left( \epsilon
\right) $ is the order-$p$ moment of ${\cal N}_{+}\left( \epsilon \right) $,
we have the following recurrence 
\begin{eqnarray*}
n_{+}^{\left( 1\right) }\left( \epsilon \right) &=&\left( 1-\epsilon \right)
^{\theta }+\theta \int_{\epsilon }^{1}v^{\theta -1}n_{+}^{\left( 1\right)
}\left( \epsilon /v\right) dv\text{ and, with }p\geq 2: \\
n_{+}^{\left( p\right) }\left( \epsilon \right) &=&\left( 1-\epsilon \right)
^{\theta }+\theta \sum_{k=1}^{p-1}\binom{p}{k}{\bf I}_{\epsilon
<1/2}\int_{\epsilon }^{1-\epsilon }v^{\theta -1}n_{+}^{\left( k\right)
}\left( \epsilon /v\right) dv \\
&&+\theta \int_{\epsilon }^{1}v^{\theta -1}n_{+}^{\left( p\right) }\left(
\epsilon /v\right) dv.
\end{eqnarray*}
Let 
\[
\widehat{n}_{+}^{\left( 1\right) }\left( q\right) :=\int_{0}^{1}\epsilon
^{q}n_{+}^{\left( 1\right) }\left( \epsilon \right) d\epsilon . 
\]
From the first recurrence on $n_{+}^{\left( 1\right) }\left( \epsilon
\right) $, recalling 
\[
\int_{0}^{1}\epsilon ^{q}\left( 1-\epsilon \right) ^{\theta }d\epsilon =%
\frac{\Gamma \left( q+1\right) \Gamma \left( \theta +1\right) }{\Gamma
\left( q+\theta +2\right) } 
\]
and using an integration by part, we obtain explicitly 
\begin{equation}
\widehat{n}_{+}^{\left( 1\right) }\left( q\right) =\frac{\Gamma \left(
q+1\right) \Gamma \left( \theta +1\right) }{\Gamma \left( q+\theta +1\right)
\left( q+1\right) },  \label{e11}
\end{equation}
leading to $n_{+}^{\left( 1\right) }\left( \epsilon \right) =\theta
\int_{\epsilon }^{1}z^{-1}\left( 1-z\right) ^{\theta -1}dz.$ The function $%
\widehat{n}_{+}^{\left( 1\right) }\left( q\right) $ has a double pole at $%
q=-1$ and $\widehat{n}_{+}^{\left( 1\right) }\left( q\right) \sim
_{q=-1}\theta \left( q+1\right) ^{-2}$, showing, as required, from
singularity analysis, that $n_{+}^{\left( 1\right) }\left( \epsilon \right)
\sim _{\epsilon \downarrow 0}-\theta \log \epsilon $.

Similarly, considering the contribution to total mass of those $\widetilde{%
\xi }_{k}$ above or below $\epsilon $, which are respectively $1_{+}\left(
\epsilon \right) :=\sum_{k\geq 1}\widetilde{\xi }_{k}{\bf I}\left( 
\widetilde{\xi }_{k}>\epsilon \right) $ and $1_{-}\left( \epsilon \right)
:=\sum_{k\geq 1}\widetilde{\xi }_{k}{\bf I}\left( \widetilde{\xi }_{k}\leq
\epsilon \right) $, it holds that 
\begin{eqnarray*}
&&1_{+}\left( \epsilon \right) \stackrel{d}{=}v_{1}{\bf I}\left(
v_{1}>\epsilon \right) +\overline{v}_{1}1_{+}^{^{\prime }}\left( \epsilon /%
\overline{v}_{1}\right) {\bf I}_{\overline{v}_{1}>\epsilon } \\
&&1_{-}\left( \epsilon \right) \stackrel{d}{=}v_{1}{\bf I}\left( v_{1}\leq
\epsilon \right) +\overline{v}_{1}1_{-}^{^{\prime }}\left( \epsilon /%
\overline{v}_{1}\right) {\bf I}_{\overline{v}_{1}>\epsilon }
\end{eqnarray*}
where $1_{\pm }^{^{\prime }}\left( .\right) $ are statistical copies of $%
1_{\pm }\left( .\right) $ respectively, independent of $\overline{v}_{1}$.
Let $x_{\pm }\left( \epsilon \right) :={\Bbb E}\left[ 1_{\pm }\left(
\epsilon \right) \right] $ stand for the mean values. Then, with 
\[
a_{+}\left( \epsilon \right) :=\theta \int_{\epsilon }^{1}v\left( 1-v\right)
^{\theta -1}dv\text{ and }a_{-}\left( \epsilon \right) :=\theta
\int_{0}^{\epsilon }v\left( 1-v\right) ^{\theta -1}dv, 
\]
we have 
\[
x_{\pm }\left( \epsilon \right) =a_{\pm }\left( \epsilon \right) +\theta
\int_{\epsilon }^{1}v^{\theta }x_{\pm }\left( \epsilon /v\right) dv. 
\]
Defining $\widehat{x}_{\pm }\left( q\right) :=\int_{0}^{1}\epsilon
^{q}x_{\pm }\left( \epsilon \right) d\epsilon $, similar computations show
that

1/ 
\begin{equation}
\widehat{x}_{+}\left( q\right) =\frac{\Gamma \left( q+1\right) \Gamma \left(
\theta +1\right) }{\Gamma \left( q+\theta +2\right) }  \label{e12}
\end{equation}
and $\widehat{x}_{+}\left( q\right) $ has a simple dominant pole at $q=-1$
with $\widehat{x}_{+}\left( q\right) \sim _{q=-1}\left( q+1\right) ^{-1}$,
showing, as required, from singularity analysis, that $x_{+}\left( \epsilon
\right) \sim _{\epsilon \downarrow 0}1$. More precisely, inverting (\ref{e12}%
), $x_{+}\left( \epsilon \right) =\left( 1-\epsilon \right) ^{\theta }$ and $%
x_{+}\left( \epsilon \right) \sim _{\epsilon \downarrow 0}1-\theta \epsilon
. $

2/ 
\begin{equation}
\widehat{x}_{-}\left( q\right) =\frac{\left( q+\theta +2\right) \Gamma
\left( \theta +1\right) }{\left( q+1\right) \left( q+2\right) }\left\{ \frac{%
\Gamma \left( q+\theta +3\right) -\Gamma \left( \theta +2\right) \Gamma
\left( q+3\right) }{\Gamma \left( \theta +2\right) \Gamma \left( q+\theta
+3\right) }\right\}  \label{e13}
\end{equation}
and $\widehat{x}_{-}\left( q\right) $ has a simple dominant pole at $q=-2$
with $\widehat{x}_{-}\left( q\right) \sim _{q=-2}\frac{\theta ^{2}}{\theta +1%
}\left( q+2\right) ^{-1}$, showing, as required, from singularity analysis,
that $x_{-}\left( \epsilon \right) \sim _{\epsilon \downarrow 0}\frac{\theta
^{2}}{\theta +1}\epsilon $.

Note that the average values $x_{\pm }\left( \epsilon \right) $ are known
explicitly through inverting $\widehat{x}_{\pm }\left( q\right) $, even if $%
\epsilon $ is not small (in particular $x_{+}\left( \epsilon \right) =\left(
1-\epsilon \right) ^{\theta }$ and $x_{-}\left( \epsilon \right) =1-\frac{%
\theta }{\theta +1}\epsilon -\left( 1-\epsilon \right) ^{\theta }$).
Singularity analysis of $\widehat{x}_{\pm }\left( q\right) $ gives the
expected small$-\epsilon $ behavior of $x_{\pm }\left( \epsilon \right) $.

Similar recurrence on higher-order moments of $1_{\pm }\left( \epsilon
\right) $ could be obtained. \newline

\begin{center}
{\bf Weighted partition: modulation}
\end{center}

Assume $\left( \widetilde{\xi }_{\left( k\right) };k\geq 1\right) $ $%
\stackrel{d}{\sim }$ GEM$\left( \theta \right) $. Let $\left( \mu _{k};k\geq
1\right) $ be a sequence of iid non-negative random variables. We shall
investigate some properties of the weighted or modulated random sequence $%
\left( \mu _{k}\widetilde{\xi }_{\left( k\right) };k\geq 1\right) $ as a new
random transformed partition of $\chi _{\mu _{1}}:=\sum_{k\geq 1}\mu _{k}%
\widetilde{\xi }_{\left( k\right) }.$ Plainly, we have 
\[
\chi _{\mu _{1}}\stackrel{d}{=}\mu _{1}v_{1}+\overline{v}_{1}\chi _{\mu
_{1}}^{\prime } 
\]
where $\chi _{\mu _{1}}^{\prime }\stackrel{d}{=}\chi _{\mu _{1}}$ is
independent of $\overline{v}_{1}\stackrel{d}{\sim }$ beta$\left( \theta
,1\right) $. Assume $c_{k}:={\Bbb E}\left[ \mu _{1}^{k}\right] <\infty $ for
all $k\geq 0$. Then, with $m_{n}:={\Bbb E}\left[ \chi _{\mu _{1}}^{n}\right] 
$, we have 
\[
m_{n}=\sum_{k=0}^{n-1}\binom{n}{k}c_{n-k}{\Bbb E}\left[ v_{1}^{n-k}\overline{%
v}_{1}^{k}\right] m_{k}+{\Bbb E}\left[ \overline{v}_{1}^{n}\right] m_{n}. 
\]
Recalling ${\Bbb E}\left[ v_{1}^{n-k}\overline{v}_{1}^{k}\right] =\theta
\left[ \left( n-k\right) !\Gamma \left( \theta +k\right) \right] /\Gamma
\left( \theta +n+1\right) $ and ${\Bbb E}\left[ \overline{v}_{1}^{n}\right]
=\theta /\left( \theta +n\right) $, setting $h_{n}:=m_{n}\Gamma \left(
\theta +n\right) /n!$, we obtain the convolution-like recurrence ($n\geq 1$) 
\[
nh_{n}=\theta \sum_{k=0}^{n-1}c_{n-k}h_{k}. 
\]
If 
\[
h\left( u\right) :=\sum_{n\geq 0}h_{n}u^{n} 
\]
is the generating function of $\left( h_{n};n\geq 1\right) $, with $c\left(
u\right) :=\sum_{n\geq 0}c_{n}u^{n}$, we obtain $uh^{\prime }\left( u\right)
+\theta h\left( u\right) =\theta c\left( u\right) h\left( u\right) $,
leading explicitly to 
\begin{equation}
h\left( u\right) =\Gamma \left( \theta \right) \exp \left\{ -\theta
\int_{0}^{u}\frac{1-c\left( v\right) }{v}dv\right\} .  \label{e14}
\end{equation}
In more details, with $\widetilde{c}_{k}:=\left( k-1\right) !c_{k}$, with $%
n\geq 1$, we finally get 
\[
m_{n}=\frac{\Gamma \left( \theta \right) n!}{\Gamma \left( \theta +n\right) }%
\sum_{k=1}^{n}\left( -\theta \right) ^{k}B_{n,k}\left( \widetilde{c}_{1},%
\widetilde{c}_{2},...,\widetilde{c}_{n-k+1}\right) 
\]
in terms of Bell polynomials.

The moments of the partition function 
\[
\chi _{\mu _{1}}\left( \beta \right) :=\sum_{k\geq 1}\left[ \mu _{k}%
\widetilde{\xi }_{\left( k\right) }\right] ^{\beta } 
\]
of $\left( \mu _{k}\widetilde{\xi }_{\left( k\right) };k\geq 1\right) $ can
be obtained similarly, observing that, with $\overline{v}_{1}^{\beta }%
\stackrel{d}{\sim }$ beta$\left( \theta /\beta ,1\right) $ 
\[
\chi _{\mu _{1}}\left( \beta \right) \stackrel{d}{=}\mu _{1}^{\beta
}v_{1}^{\beta }+\overline{v}_{1}^{\beta }\chi _{\mu _{1}}\left( \beta
\right) . 
\]
\newline

{\em Example:}

When $\left( \mu _{k};k\geq 1\right) $ is a sequence of iid non-negative
random variables drawn for Bernoulli$\left( p\right) $ distribution, with $%
c_{k}=p$, $k\geq 1$, one can check that 
\[
m_{n}=\frac{\Gamma \left( p\theta +n\right) \Gamma \left( \theta \right) }{%
\Gamma \left( p\theta \right) \Gamma \left( \theta +n\right) },\text{ }n\geq
1 
\]
solves the recurrence. This shows that $\chi _{\mu _{1}}:=\sum_{k\geq 1}\mu
_{k}\widetilde{\xi }_{\left( k\right) }$ has beta$\left( p\theta ,\left(
1-p\right) \theta \right) $ distribution in this case (see \cite{Hirt}). $%
\square $\newline

\begin{center}
{\bf Structural Measure}
\end{center}

With $\beta >\beta _{*}=0$, we also have in this case 
\begin{equation}
\widetilde{\phi }\left( \beta \right) =\frac{\Gamma \left( \beta \right)
\Gamma \left( \theta +1\right) }{\Gamma \left( \theta +\beta \right) }%
=\int_{0}^{1}x^{\beta }\frac{\theta }{x}\left( 1-x\right) ^{\theta -1}dx.
\label{eq37}
\end{equation}
The measure $\sigma \left( dx\right) :=\frac{\theta }{x}\left( 1-x\right)
^{\theta -1}{\bf I}_{x\in \left( 0,1\right) }dx$ is the structural measure
of the GEM$\left( \theta \right) $ partition.\newline

\begin{center}
{\bf Typical Fragment Size }$\widetilde{\xi }${\bf \ from} $\left( 
\widetilde{\xi }_{k},k\geq 1\right) $
\end{center}

Its moment function is given by 
\[
\varphi _{\widetilde{\xi }}\left( q\right) =\sum_{k\geq 1}{\Bbb E}\left( 
\widetilde{\xi }_{k}\right) \varphi _{\widetilde{\xi }_{k}}\left( q\right) 
\]
where $\varphi _{\widetilde{\xi }_{k}}\left( q\right) :={\Bbb E}\widetilde{%
\xi }_{k}^{q}=\frac{\Gamma \left( q+1\right) \Gamma \left( \theta +1\right) 
}{\Gamma \left( \theta +q+1\right) }\left( \frac{\theta }{\theta +q}\right)
^{k-1}$ is the moment function of $\widetilde{\xi }_{k}.$ As a result 
\[
\varphi _{\widetilde{\xi }}\left( q\right) =\frac{\Gamma \left( q+1\right)
\Gamma \left( \theta +1\right) }{\Gamma \left( \theta +q+1\right) \left(
\theta +1\right) }\sum_{k\geq 1}\left( \frac{\theta ^{2}}{\left( \theta
+1\right) \left( \theta +q\right) }\right) ^{k-1} 
\]
\begin{equation}
=\frac{\Gamma \left( q+1\right) \Gamma \left( \theta +1\right) }{\Gamma
\left( \theta +q+1\right) }\frac{\theta +q}{\theta +\left( \theta +1\right) q%
}  \label{eq38}
\end{equation}
showing that $\widetilde{\xi }\stackrel{d}{=}\widetilde{\eta }\cdot R$ where 
$\widetilde{\eta }\stackrel{d}{\sim }$ beta$\left( 1,\theta \right) $ is
independent of the $\left[ 0,1\right] -$valued random variable $R$, with
moment function ${\Bbb E}R^{q}=$ $\frac{\theta +q}{\theta +\left( \theta
+1\right) q}$. This interprets as follows: let $B$ be a Bernoulli random
variable with parameter $\frac{1}{\theta +1}$ and $C\stackrel{d}{\sim }$ beta%
$\left( \frac{\theta }{\theta +1},1\right) $ a random variable on $\left[
0,1\right] $, independent of $B$. Then, $R$ is a $\left[ 0,1\right] $-valued
random variable satisfying 
\[
R\stackrel{d}{=}B+\left( 1-B\right) \cdot C 
\]
Indeed, 
\[
{\Bbb E}R^{q}=\frac{1}{\theta +1}+\frac{\theta }{\theta +1}\frac{\theta
/\left( \theta +1\right) }{\theta /\left( \theta +1\right) +q}=\frac{\theta
+q}{\theta +\left( \theta +1\right) q}. 
\]
\newline

\begin{center}
{\bf Size-biased Picking }$\widetilde{\eta }${\bf \ from} $\left( \widetilde{%
\xi }_{k},k\geq 1\right) $
\end{center}

Its moment function is 
\begin{equation}
{\Bbb E}\widetilde{\eta }^{q}:={\Bbb E}\sum_{k\geq 1}\widetilde{\xi }%
_{k}^{q+1}=\widetilde{\phi }\left( q+1\right) =\frac{\Gamma \left(
q+1\right) \Gamma \left( \theta +1\right) }{\Gamma \left( \theta +q+1\right) 
}  \label{eq39}
\end{equation}
which is the moment function of a beta$\left( 1,\theta \right) $ distributed
random variable (as required from the size-biased picking invariance
property of $\left( \widetilde{\xi }_{k},k\geq 1\right) $).

The waiting time paradox 
\[
\widetilde{\eta }\succeq _{st}\widetilde{\xi } 
\]
is clearly satisfied from the decomposition $\widetilde{\xi }\stackrel{d}{=}%
\widetilde{\eta }\cdot R$.\newline

In the next two examples, computations on spacings are quite involved. We
skip them, focusing on the simplest aspects.\newline

\subsection{Two Additional Examples}

\indent

We briefly sketch some properties of related partition models.\newline

$\bullet $ Let $\alpha \in \left( 0,1\right) $ and put $\overline{\alpha }%
:=1-\alpha $. Assume $\Pi \left( dx\right) =\theta \overline{\alpha }%
x^{-\left( 1+\alpha \right) }{\bf I}_{x\in \left( 0,1\right) }dx.$ Then,
with $a=\frac{\theta \overline{\alpha }}{\alpha }>0$, 
\[
\overline{\Pi }\left( x\right) =a\left( x^{-\alpha }-1\right) \text{ and }%
\overline{\Pi }^{-1}\left( s\right) =\left( 1+s/a\right) ^{-1/\alpha }. 
\]
We have $\xi _{\left( k\right) }=\left( 1+S_{k}/a\right) ^{-1/\alpha }$ and 
\[
{\Bbb E}\sum_{k\geq 1}\xi _{\left( k\right) }^{\beta }=\int_{0}^{1}x^{\beta }%
\frac{\theta \overline{\alpha }}{x^{1+\alpha }}dx=\frac{\theta \overline{%
\alpha }}{\beta -\alpha }, 
\]
for $\beta >\alpha =\beta _{*}>0.$ Note that $\frac{a}{k}\left( \xi _{\left(
k\right) }^{-\alpha }-1\right) \rightarrow 1$ almost surely as $k\uparrow
\infty $ and $\xi _{\left( k\right) }\sim \left( \frac{k}{a}\right)
^{-1/\alpha }$ goes to $0$ algebraically fast with $k$ (like $k^{-1/\alpha }$%
) in this case. Spacings are more involved.\newline

$\bullet $ Let $\alpha \in \left( 0,1\right) $, put $\overline{\alpha }%
:=1-\alpha $ and assume 
\[
\Pi \left( dx\right) =\frac{\theta }{\Gamma \left( \alpha \right) \Gamma
\left( \overline{\alpha }\right) }x^{-\left( 1+\alpha \right) }\left(
1-x\right) ^{\alpha -1}{\bf I}_{x\in \left( 0,1\right) }dx 
\]
with $\int_{0}^{1}x\Pi \left( dx\right) =\theta .$ We have 
\[
{\Bbb E}\sum_{k\geq 1}\xi _{\left( k\right) }^{\beta }=\frac{\theta }{\Gamma
\left( \alpha \right) \Gamma \left( \overline{\alpha }\right) }%
\int_{0}^{1}x^{\left( \beta -\alpha \right) -1}\left( 1-x\right) ^{\alpha
-1}dx=\frac{\theta \Gamma \left( \beta -\alpha \right) }{\Gamma \left( 
\overline{\alpha }\right) \Gamma \left( \beta \right) }, 
\]
for $\beta >\alpha =\beta _{*}>0.$

When $\alpha =1/2$, $\overline{\Pi }\left( x\right) =\frac{2\theta }{\pi }%
\left( \frac{1-x}{x}\right) ^{1/2}$ and $\overline{\Pi }^{-1}\left( s\right)
=\left( 1+\left( \frac{\pi s}{2\theta }\right) ^{2}\right) ^{-1}$. We have $%
\xi _{\left( k\right) }=\left( 1+\left( \frac{\pi S_{k}}{2\theta }\right)
^{2}\right) ^{-1}$and $\frac{2\theta }{\pi k}\left( \xi _{\left( k\right)
}^{-1}-1\right) ^{1/2}\rightarrow 1$ almost surely as $k\uparrow \infty $.
As a result, $\xi _{\left( k\right) }\sim \left( \frac{2\theta }{\pi k}%
\right) ^{2}$ goes slowly to $0$ (like $k^{-2}$) in this case. The
associated size-biased picking random variable $\eta $ follows the Arcsine$%
\left( 1/2\right) $ law. \newline

In both examples, an $\alpha $ larger than $1$ would violate the condition
that $\Pi $ is a L\'{e}vy measure. \newline

\section{The Bounded Partition Model with a Poissonian Number of Fragments}

\indent

We finally briefly show that the partitioning model based on L\'{e}vy
measure concentrated on $\left( 0,1\right) $ is also of some statistical
relevance in the case of a bounded L\'{e}vy measure for jumps. This model
does not seem to have received attention in the literature. \newline

In the bounded case, let $\mu :=\overline{\Pi }\left( 0\right) <\infty $. In
this case 
\begin{equation}
{\Bbb E}e^{-\lambda \chi }=\exp \left\{ -\mu \left(
1-\int_{0}^{1}e^{-\lambda x}F\left( dx\right) \right) \right\}  \label{eq40}
\end{equation}
where $F\left( dx\right) =\Pi \left( dx\right) /\mu $ is a probability
distribution with mean value $\theta /\mu <1$.\newline

\subsection{Poisson Partition of $\chi $: the Model}

\indent

Hence, with $u_{k},$ $k\geq 1$ an iid $\left( 0,1\right) -$valued uniform
sequence, $P_{\mu }$ a Poisson random variable with intensity $\mu $, $\chi
=\sum_{k=1}^{P_{\mu }}\overline{F}^{-1}\left( u_{k}\right) $ belongs to the
class of compound Poisson random variables. Stated differently 
\begin{equation}
\chi \stackrel{d}{=}\sum_{k=1}^{P_{\mu }}\overline{F}^{-1}\left( u_{\left(
k\right) ,P_{\mu }}\right)  \label{eq41}
\end{equation}
where $u_{\left( 1\right) ,P_{\mu }}<..<u_{\left( P_{\mu }\right) ,P_{\mu }}$
is obtained from sample $u_{1},..,u_{P_{\mu }}$ while ordering the
constitutive terms. We note that $\chi $ has an atom at $\chi =0$ with
probability $e^{-\mu }$ and that there are finitely many (Poissonian)
fragments. This partition is also $\chi \stackrel{d}{=}\sum_{k=1}^{P_{\mu }}%
\overline{\Pi }^{-1}\left( \mu u_{\left( k\right) ,P_{\mu }}\right) $ where,
when $\mu \uparrow \infty $, $P_{\mu }\stackrel{a.s.}{\rightarrow }\infty $
and $\left( \mu u_{\left( 1\right) ,P_{\mu }},..,\mu u_{\left( k\right)
,P_{\mu }}\right) \stackrel{d}{\rightarrow }\left( S_{1},..,S_{k},..\right) $
a Poisson point process on ${\Bbb R}^{+}.$

Defining $\xi _{\left( k\right) ,P_{\mu }}:=\overline{F}^{-1}\left(
u_{\left( k\right) ,P_{\mu }}\right) $ , with $\xi _{\left( 1\right) ,P_{\mu
}}>..>\xi _{\left( P_{\mu }\right) ,P_{\mu }}$, we get similarly 
\begin{equation}
\phi \left( \beta \right) ={\Bbb E}\sum_{k=1}^{P_{\mu }}\xi _{\left(
k\right) ,P_{\mu }}^{\beta }=\int_{0}^{1}x^{\beta }\Pi \left( dx\right)
\label{eq42}
\end{equation}
and the structural measure is $\Pi \left( dx\right) =\mu F\left( dx\right) $
which is bounded with $\int_{0}^{1}x\Pi \left( dx\right) =\theta $.\newline

{\em Remark:} Assume that the distribution $F$ is the one of a beta$\left(
\theta /\left( \mu -\theta \right) ,1\right) $, $\mu >\theta $, with mean
value $\theta /\mu $. Then 
\[
\overline{\Pi }\left( x\right) :=\mu \overline{F}\left( x\right) =\mu \left(
1-x^{\theta /\left( \mu -\theta \right) }\right) \rightarrow _{\mu \uparrow
\infty }-\theta \log x 
\]
which is the L\'{e}vy-measure tail of the limiting Dickman model discussed
in Section 4.1. $\square $\newline

\begin{center}
{\bf Typical Fragment Size from} $\left( \xi _{\left( k\right) ,P_{\mu
}},k=1,..,P_{\mu }\right) $
\end{center}

In this case, the typical fragment size from $\left( \xi _{\left( k\right)
,P_{\mu }},k=1,..,P_{\mu }\right) $ clearly is the $\left( 0,1\right) -$%
valued random variable, say $\xi $, with probability distribution $F_{\xi
}\left( x\right) =F\left( x\right) $ with ${\Bbb E}\xi =\theta /\mu .$%
\newline

\begin{center}
{\bf Size-biased Picking from} $\left( \xi _{\left( k\right) ,P_{\mu
}},k=1,..,P_{\mu }\right) $
\end{center}

Let $\eta $ be a $\left( 0,1\right) -$valued random variable taking the
value $\xi _{\left( k\right) ,P_{\mu }}$ with probability $\frac{1}{\theta }%
\xi _{\left( k\right) ,P_{\mu }}$ given $\left( \xi _{\left( k\right)
,P_{\mu }},k=1,..,P_{\mu }\right) $. This random variable corresponds to a
size-biased picking from $\left( \xi _{\left( k\right) ,P_{\mu
}},k=1,..,P_{\mu }\right) $. Its moment function is 
\begin{equation}
\varphi _{\eta }\left( q\right) ={\Bbb E}\eta ^{q}:={\Bbb E}\frac{1}{\theta }%
\sum_{k=1}^{P_{\mu }}\xi _{\left( k\right) ,P_{\mu }}^{q+1}=\frac{1}{\theta }%
\phi \left( q+1\right)  \label{eq43}
\end{equation}
and $\eta $ has probability distribution $F_{\eta }\left( x\right) =\frac{%
\mu }{\theta }\int_{0}^{x}zF_{\xi }\left( dz\right) .$

The waiting time paradox reads 
\[
\eta \succeq _{st}\xi , 
\]
since $F_{\eta }\left( x\right) \leq $ $F_{\xi }\left( x\right) $ for all $%
x\in \left[ 0,1\right] .$

\subsection{Spacings}

\indent

As in the unbounded case, spacings of the Poisson partition of $\chi $
deserve interest.

Defining the spacings $\widetilde{\xi }_{\left( k\right) ,P_{\mu }}:=\xi
_{\left( k-1\right) ,P_{\mu }}-\xi _{\left( k\right) ,P_{\mu }}$ (with $\xi
_{\left( 0\right) ,P_{\mu }}:=1$ and $\xi _{\left( P_{\mu }+1\right) ,P_{\mu
}}:=0$), $k=1,..,P_{\mu }+1$, then, $\left( \widetilde{\xi }_{\left(
k\right) ,P_{\mu }},k=1,..,P_{\mu }+1\right) $ constitutes a new sequence of 
$\left( 0,1\right) -$valued random variables with clearly $%
\sum_{k=1}^{P_{\mu }+1}\widetilde{\xi }_{\left( k\right) ,P_{\mu }}=1.$ Let
for example 
\[
\widetilde{\phi }\left( \beta \right) :={\Bbb E}\sum_{k=1}^{P_{\mu }+1}%
\widetilde{\xi }_{\left( k\right) ,P_{\mu }}^{\beta }. 
\]
With $u_{\left( 0\right) ,P_{\mu }}:=0$ and $u_{\left( P_{\mu }+1\right)
,P_{\mu }}:=1$, it can be obtained in general from 
\begin{equation}
\widetilde{\phi }\left( \beta \right) ={\Bbb E}\sum_{k=1}^{P_{\mu }+1}\left[
\left( \overline{F}^{-1}\left( u_{\left( k-1\right) ,P_{\mu }}\right) -%
\overline{F}^{-1}\left( u_{\left( k\right) ,P_{\mu }}\right) \right) ^{\beta
}\right]  \label{eq44}
\end{equation}
so that, given $P_{\mu }=p$, the joint law of $\left( u_{\left( k-1\right)
,p};u_{\left( k\right) ,p}\right) $ is needed to compute $\widetilde{\phi }%
\left( \beta \right) $ or the structural measure $\sigma $ such that $%
\widetilde{\phi }\left( \beta \right) =\int_{0}^{1}x^{\beta }\sigma \left(
dx\right) $.\newline

{\em Example:}

Let $\Pi \left( dx\right) =\mu {\bf I}_{x\in \left( 0,1\right) }dx.$ Then $%
F\left( dx\right) $ is the uniform distribution ($\overline{F}^{-1}\left(
x\right) =1-x$) and $\theta =\mu /2.$ We obtain ${\Bbb E}\sum_{k=1}^{P_{\mu
}}\xi _{\left( k\right) ,P_{\mu }}^{\beta }=\frac{2\theta }{\beta +1}=\phi
\left( \beta \right) .$ The random variable $\xi $ is uniform and the
size-biased picking random variable $\eta $ has distribution beta$\left(
2,1\right) $.

Concerning spacings, we have 
\[
\widetilde{\phi }\left( \beta \right) ={\Bbb E}\sum_{k=1}^{P_{\mu }+1}\left(
\xi _{\left( k-1\right) ,P_{\mu }}-\xi _{\left( k\right) ,P_{\mu }}\right)
^{\beta }=\int_{0}^{1}x^{\beta }\sigma \left( dx\right) . 
\]
The tail of the structural measure for spacings reads 
\begin{eqnarray*}
\int_{x}^{1}\sigma \left( dz\right) &=&:\overline{\sigma }\left( x\right) =%
{\Bbb E}\sum_{k=1}^{P_{\mu }+1}{\Bbb P}\left( \xi _{\left( k-1\right)
,P_{\mu }}-\xi _{\left( k\right) ,P_{\mu }}>x\right) \\
&=&\sum_{p\geq 0}\frac{e^{-\mu }\mu ^{p}}{p!}\left( p+1\right) \left(
1-x\right) ^{p}=e^{-\mu x}\left( 1+\mu \left( 1-x\right) \right)
\end{eqnarray*}
recalling that $\overline{F}_{\widetilde{\xi }_{\left( k\right) ,:p}}\left(
x\right) =\left( 1-x\right) ^{p}$ is the tail distribution function of
uniform spacings $\widetilde{\xi }_{\left( k\right) ,p}:=\xi _{\left(
k-1\right) ,p}-\xi _{\left( k\right) ,p}$ (with $\xi _{\left( 0\right)
,p}:=1 $ and $\xi _{\left( p+1\right) ,p}:=0$), for any $k=1,..,p+1.$ $%
\square $

\end{document}